\documentclass[structabstract]{aa}
\usepackage{graphicx}
\usepackage{natbib}
\usepackage[dvipsnames]{xcolor}
\usepackage{hyperref}
\usepackage{txfonts}
\usepackage{longtable}

\def\color#1{}

\title{2021 occultations and transits of Linus orbiting (22) Kalliope: I.~Polygonal and `cliptracing' algorithms}
\titlerunning{2021 occultations and transits of Linus orbiting (22) Kalliope}

\author{
    M.~Bro\v{z}\inst{\ref{prague}}         \and 
    J.~\v{D}urech\inst{\ref{prague}}       \and 
    M.~Ferrais\inst{\ref{arecibo}}         \and 
    H.-J.~Lee\inst{\ref{kasi}}             \and 
    M.-J.~Kim\inst{\ref{kasi}}             \and 
    D.-G.~Roh\inst{\ref{kasi}}             \and 
    H.-S.~Yim\inst{\ref{kasi}}             \and 
    E.~Jehin\inst{\ref{liege}}             \and 
    A.~Burdanov\inst{\ref{mit}}            \and 
    J.~de~Wit\inst{\ref{mit}}              \and 
    P.~Fatka\inst{\ref{asu}}               \and 
    J.~Hanu\v{s}\inst{\ref{prague}}        \and 
    B.~Carry\inst{\ref{nice}}                   
}

\institute{
     Charles University, Faculty of Mathematics and Physics, Institute of Astronomy, V~Hole{\v s}ovi{\v c}k{\'a}ch 2, 18000 Prague, Czech Republic%
     \label{prague}%
     \and 
     Arecibo Observatory, University of Central Florida, HC-3 Box 53995, Arecibo, PR 00612, USA
     \label{arecibo}
     \and 
     Korea Astronomy and Space Science Institute, 776 Daedeok-daero, Yuseong-gu, Daejeon, South Korea%
     \label{kasi}
     \and 
     Space sciences, Technologies and Astrophysics Research (STAR) Institute, University of Liège, Allée du 6 Août 19, 4000 Liège, Belgium
     \label{liege}
     \and 
     Department of Earth, Atmospheric and Planetary Sciences, MIT, 77 Massachusetts Avenue, Cambridge, MA 02139, USA
     \label{mit}
     \and 
     Academy of Sciences of the Czech Republic, Astronomical Institute, Fri{\v c}ova 1, 25165 Ond\v rejov, Czech Republic%
     \label{asu}
     \and 
     Université Côte d'Azur, Observatoire de la Côte d'Azur, CNRS, Laboratoire Lagrange, France
     \label{nice}
}

\date{Received x-x-2023 / Accepted x-x-2023}

\abstract
   {}
   {
The satellite Linus orbiting the main-belt asteroid (22) Kalliope
exhibited mutual occultation and transit events in late 2021.
A photometric campaign was organized and observations were taken
by the TRAPPIST-South, SPECULOOS-Artemis, OWL-Net, and BOAO telescopes,
with the goal to further constrain dynamical and photometric models
of this sizeable asteroid--satellite system.
   } 
   {
Our dynamical model is sufficiently complex,
with multipoles (up to the order $\ell = 2$),
internal tides,
and external tides.
The model was constrained by
astrometry (spanning 2001--2021),
occultations,
adaptive-optics imaging,
calibrated photometry,
as well as relative photometry.
Our photometric model was substantially improved.
A new precise (${<}\,0.1\,{\rm mmag}$) light curve algorithm was implemented,
based on polygon intersections,
which are computed exactly ---
by including partial eclipses and partial visibility of polygons.
Moreover, we implemented a `cliptracing' algorithm,
based again on polygon intersections,
in which partial contributions to individual pixels
are computed exactly.
Both synthetic light curves and synthetic images are then very smooth.
   }
   {
Based on our combined solution,
we confirmed the size of Linus,
$(28\pm 1)\,{\rm km}$.
However, this solution exhibits some tension
between the light curves
and the PISCO speckle-interferometry dataset,
acquired contemporarily with the 2021 events.
This indicates that improvements of the shape are still possible.
In most solutions, Linus is darker than Kalliope,
with the single-scattering albedos $A_{\rm w} = 0.40$ vs. $0.44$.
This is confirmed on deconvolved images.
A~detailed revision of astrometric data allowed us
to revise also the $J_2 \equiv -C_{20}$ value of Kalliope.
Most importantly, a~homogeneous body is excluded.
For a differentiated body, two solutions exist:
low-oblateness ($C_{20} \simeq -0.12$),
with a~spherical iron core,
and alternatively, high-oblateness ($C_{20} \simeq -0.22$)
with an elongated iron core.
These correspond to the low- and high-energy collisions, respectively,
studied by means of SPH simulations in our previous work.
   }
   {}
 

\begin{document}

\maketitle


\section{Introduction}

Mutual events between asteroids and their satellites
are not rare
(e.g., \citealt{Pravec_1997Icar..127..431P,Ragozzine_2009AJ....137.4766R,Wong_2019AJ....157..203W,Scheirich_2022PSJ.....3..163S}).
The timings of occultations, transits, or eclipses
can be used for various precise measurements.
Recently, such timings were used to measure the outcome
of the DART experiment
\citep{Cheng_2018P&SS..157..104C,Statler_2022PSJ.....3..244S,Thomas_2023}.

The (22) Kalliope and Linus binary system exhibited the eclipse events back in 2007
\citep{Descamps_2008Icar..196..578D}.
With the shape of Kalliope derived from light curves,
and an assumed spherical shape of Linus,
they obtained its size $(28\pm 2)\,{\rm km}$,
based on magnitude drops due to eclipses.
This value is compatible with the shadow of Linus,
which was observed during the stellar occultation
event on Nov 7th 2006.

Here we use more complex dynamical and photometric models
of the (22) Kalliope and Linus system
to interpret 2021 mutual occultation and transit events.
At the same time, our preferred shape model
was constrained by the VLT/SPHERE high-resolution AO imaging
\citep{Ferrais_2022A&A...662A..71F}.
Last but not least, a new context has been set up
by the discovery of the first M-type Kalliope family
\citep{Broz_2022A&A...664A..69B},
which strongly suggests a~differentiated interior.

The paper is organized as follows.
In Sec.~\ref{data}, new light curve data are presented,
together with other data used to constrain the model.
In Sec.~\ref{polygon}, a new polygonal light curve algorithm is described.
It is generally needed to achieve high precision (${<}\,0.1\,{\rm mmag}$),
whenever a moon is relatively small,
or signal to noise is relatively high.
In Sec.~\ref{cliptracing}, a `cliptracing' is described,
which is used to compute as-smooth-as-possible synthetic images.
Additionally, in Sec.~\ref{occultation}, a stellar occultation algorithm
is also explained.
In Sec.~\ref{results}, results of our astrometric and photometric models
are presented.


\section{Observational data}\label{data}

\subsection{New light curves}

We obtained a calibrated photometry in the Rc band at the 0.6-m TRAPPIST-South telescope
\citep{Jehin_2011Msngr.145....2J}
and in the custom `z cut' filter at the 1-m Artemis telescope
\citep{Burdanov_2022PASP..134j5001B}
of the SPECULOOS network
\citep{Delrez_2018SPIE10700E..1ID}.
The narrow-band `z cut' filter
(transmittance ${>}90\%$ from 860\,nm to 1100\,nm)
was used to avoid saturation of the CCD pixels
and to suppress the effect of atmospheric water absorption.
The relative precision of these data is about $3\,{\rm mmag}$.
Additional offsets are present between individual nights,
which cannot be explained by the variable distance or the phase curve.
Hence, the absolute precision is (at worst) $80\,{\rm mmag}$.


Out of all light curves,
the first one includes a total occultation of Linus.
The second is a total transit of Linus.
The third is a partial transit of Linus,
when only a dark part of (22) was hidden.
The fourth is a partial occultation of Linus,
when approximately half of Linus was hidden.
The event times were prediced using the ephemeris from \cite{Ferrais_2022A&A...662A..71F}.


We also obtained a relative photometry in the Rc band
at the 1.6-m OWL-Net
\citep{Park_2018AdSpR..62..152P}
and 1.8-m BOAO
\citep{Sung_2012PKAS...27...95S}
telescopes.
The data from
2459557,
2459559
were removed,
because signal to noise was worse due to weather conditions.

Additional dense light curves were taken from the DAMIT database
\citep{Durech_2010A&A...513A..46D}, in particular
2454175,
2455965,
and additional reference light curve from
2459711,
was obtained by the BlueEye600 telescope
\citep{Durech_2018Icar..304..101D}
to constrain the rotation phase.

Light curves from the previous series of events
\citep{Descamps_2008Icar..196..578D}
was also used, in particular
2454167,
2454176,
(denoted as `16', `25' by \citealt{Descamps_2008Icar..196..578D}),
together with their reference curves
(`14', `27').
They include a total eclipse of Linus,
and an annular eclipse of Kalliope.
These data were precisely digitized from figures.
All data were consistently converted from the UTC to the TDB time scale.
The summary of observational circumstances is presented in Tab.~\ref{tab1}.

\subsection{Calibrated photometry}

Moreover, we used sparse calibrated photometry from Gaia
\citep{Brown_2018A&A...616A...1G},
namely 19 points,
with a $<1\,{\rm mmag}$ precision.
They were transformed from G to V as follows
\citep{Leeuwen_2018gdr2.reptE....V}:
\begin{equation}
G-V = a + b(B-V) + c(B-V)^2 + d(B-V)^3\,,
\end{equation}
where
$a = -0.02907$,
$b = -0.02385$,
$c = -0.22970$,
$d = -0.001768$,
and
$B-V = 0.70\,{\rm mag}$ was taken from \cite{Lupishko_1982AVest..16..101L}.
In order to interpret these high-precision data,
a precise shape model of (22) is necessary;
otherwise the phase curve could not be fitted at all.
The data from
2457865.8993,
2457886.9754
were removed from the fit,
because they were too offset with respect to the neighbouring points.


In order to extend the phase coverage,
from $2$ up to $20^\circ$,
we also included the calibrated UBV photometry from
\citet{Gehrels_1962ApJ...135..906G},
\citet{Scaltriti_1978Icar...34...93S},
\citet{Surdej_1986A&A...170..167S}.

\subsection{Astrometry}

All astrometric data were summarized in \cite{Ferrais_2022A&A...662A..71F}.
In this work,
SCAM, PHARO, IRCAL, NIRC2, NACO, NIRI, BTA, SOR, SPHERE, and PISCO datasets
were used.
Out of the Keck/NIRC2 dataset,
four close-in-time measurements,
2452270,
2452270,
2452270,
2452270,
were removed from the fit,
because they exhibited a systematic photocentre offset.
We checked the original images,
which were fuzzy due to tracking problems.
(Actually, two Linuses were present on one image!)

The astrometry of \citet{Descamps_2008Icar..196..578D},
inferred from the eclipse events,
2454167,
2454167,
2454176,
2454176,
was removed due to unrealistic uncertainties,
which were certainly correlated with the shape.
Yet, the shape was somewhat uncertain at that time
(cf. their Fig.~2 and \citealt{Ferrais_2022A&A...662A..71F}).

The C2PU/PISCO dataset \citep{Scardia_2019AN....340..771S},
measured by speckle interferometry,
is important because it is temporarily close to the 2021 events.
The data from
2459580.6281
was removed due to a substantial offset with respect to
the neighbouring points.

Photocentre-to-centre-of-mass corrections
were derived from our photometric model (Sec.~\ref{polygon}),
and applied consistently to all astrometric data.

\subsection{AO imaging}

The AO imaging by the VLT/SPHERE/ZIMPOL was already described in
\citet{Ferrais_2022A&A...662A..71F}.
In this work, 35 deconvolved images were used to obtain observed silhouettes
and to constrain the orientation of (22),
or to prevent unwanted pole orientations.
The uncertainty of silhouettes was nominally assumed $1\,{\rm mas}$;
the pixel scale is $3.6\,{\rm mas}\,{\rm pxl}^{-1}$,
so it corresponds to a sub-pixel precision.
It is a very useful regularisation of our dynamical model,
because it is sensitive to the inclination of Linus' orbit
with respect to the equator of (22).

Moreover, some of the deconvolved images include Linus itself,
in the limited field of view. This can be used to constrain
the albedos of both bodies, or whether Linus is darker (or brighter)
than (22) Kalliope.


\subsection{Stellar occultations}

The occultation of (22) Kalliope was already used to refine
the shape model \citep{Ferrais_2022A&A...662A..71F}.
Here we use the most precise astrometric position
inferred from the Nov 7th 2006 occultation,
which included also Linus \citep{Descamps_2008Icar..196..578D}.
In this case, the astrometry is free from any photocentre offsets.

Another Mar 2nd 2022 occultation did not include Linus, unfortunately.
For (22) Kalliope, only minor systematic differences at the limb
were apparent, with respect to the nominal shape model.

\subsection{Shape model}

In this work, we use the ADAM shape model from our previous work
\citep{Ferrais_2022A&A...662A..71F}.
It was well-constrained by the AO imaging, light curves, and occultations.
However, it could be revised if the pole inferred from dynamics
is significantly different from the nominal pole
($l = 195^\circ$, $b = 4^\circ$).
For a homogeneous body, the oblateness is $J_2 = -C_{20} = 0.1586$.

This shape was derived with a regularisation of the centre of mass
and the moment of inertia tensor, in order to enforce a rotation
about the principal axis. The respective photometry is thus
insensitive to any offsets of the centre of mass;
it is like with re-centring.

However, there are possible offsets due to the inhomogeneous structure.
Hydrodynamic simulations of collisions with a differentiated body
described in \cite{Broz_2022A&A...664A..69B},
suggest mantle ejection, core deformation, gravitational reaccumulation,
and asymmetric deposition (cf. 2~hills, elongated core).
Consequently, centre-of-mass offsets will be eventually treated
as free parameters.

\subsection{Scattering law}

Initially, we assumed the Hapke scattering law \citep{Hapke_1981JGR....86.3039H},
with parameters similar as for (216) Kleopatra,
which is also also an M-type
\citep{Descamps_2008Icar..196..578D}.
Namely, the opposition effect amplitude $B_0 = 1.276$,
the o. e. width $h = 0.0470$,
the asymmetry factor $g = -0.254$,
the roughness $\bar\theta = 20^\circ$.

However, preliminary tests showed that the roughness
must be $\bar\theta < 20^\circ$,
otherwise the light curve is too curved.
We thus preferred roughness from \cite{Spjuth_2009PhDT.......588S},
their Tab.~6.3,
for (2867) \v Steins,
$\bar\theta = (11\pm 1)^\circ$.

For the phase curve fitting,
we also need the spectral slope $\gamma = 0.45$ between V, Rc bands.
This value was derived from the observed spectra of (22) Kalliope
\citep{Demeo_2009Icar..202..160D}.


\begin{table}
\caption{Observational circumstances of photometric data.}
\label{tab1}
\centering
\footnotesize
\begin{tabular}{ll@{\kern.2cm}l@{\kern.2cm}l@{\kern.2cm}l}
\hline
\vrule width 0pt height 9pt
Time & Set & Event & Filter & Reference \\
\hline
\vrule width 0pt height 8pt
2454165.5 &    & ref.		& R     & \cite{Descamps_2008Icar..196..578D} \\
2454167.5 &    & ecl. of L.	& R     & \cite{Descamps_2008Icar..196..578D} \\
2454175.3 &    & ref.           & R     & \cite{Hanus_2016A_A...586A.108H} \\
2454176.5 &    & ecl. of (22)	& R     & \cite{Descamps_2008Icar..196..578D} \\
2454177.5 &    & ref.		& R     & \cite{Descamps_2008Icar..196..578D} \\
2455965.4 &    & ref.           & R     & \cite{Vernazza_2021A_A...654A..56V} \\
2459546.0 &  9 & ref.		& Rc    & {\tiny BOAO} \\
2459546.6 & 13 & occ. of L.	& Rc    & {\tiny OWL-Net, Mt. Lemmon} \\
2459546.9 &  1 & occ. of L.	& Rc    & {\tiny TRAPPIST-South} \\
2459547.1 & 10 & ref.		& Rc    & {\tiny BOAO} \\
2459547.7 &  2 & ref.		& Rc    & {\tiny TRAPPIST-South} \\
2459548.5 &  3 & tra. of L.	& z cut & {\tiny SPECULOOS-Artemis} \\
2459548.6 &  4 & ref.		& z cut & {\tiny SPECULOOS-Artemis} \\
2459551.9 & 15 & tra. of L.	& Rc    & {\tiny OWL-Net, Mt. Bohyun} \\
2459553.6 & 14 & occ. of L.	& Rc    & {\tiny OWL-Net, Mt. Lemmon} \\
2459555.6 &  5 & tra. of L.	& Rc    & {\tiny TRAPPIST-South} \\
2459556.7 &  6 & ref.		& Rc    & {\tiny TRAPPIST-South} \\
2459557.4 &  7 & occ. of L.	& z cut & {\tiny SPECULOOS-Artemis} \\
2459559.3 &  8 & tra. of L.	& z cut & {\tiny SPECULOOS-Artemis} \\
2459711.3 &    & ref.           & Rc    & {\tiny BE600} \\
\hline
\end{tabular}
\tablefoot{
Time corresponds to an approximate beginning of observation,
`set' is the dataset number (for reference),
`ref.' a reference light curve,
`tra.' a transit,
`occ.' an occultation.
}
\end{table}


\section{Polygonal light curve algorithm}\label{polygon}

First, we implemented a new polygonal light curve algorithm
to compute light curves of asteroid--satellite systems
as precisely as possible.
Apart from a stand-alone Fortran module,
it was included in our asteroid modelling tool {\tt Xitau}%
\footnote{\url{https://sirrah.troja.mff.cuni.cz/~mira/xitau/}},
The algorithm is based on an analytical computation of polygon intersections \citep{Vatti_1992}
and the Clipper2 C++ library%
\footnote{\url{https://github.com/AngusJohnson/Clipper2}}.
This is a similar approach as in the stellar modelling tool {\tt Phoebe2}
\citep{Prsa_2016ApJS..227...29P},
but complicated by the fact that we have to compute not only the visibility,
but also non-convex shadowing, which is critical for asteroids.

In our model,
everything is orbiting, rotating, or being affected by free parameters,
including the Sun, Earth, (22), Linus.
At every time step,
occultations,
transits,
or eclipses
must be computed efficiently and exactly.
This includes not only total, but also partial or annular events.
Moreover, the algorithm should work even if
polygons of the 1st body are several times smaller that those of the 2nd body.
Of course, uncertainties of the shape itself cannot be avoided
(e.g., sphere vs. icosahedron),
but the discretisation errors due to finite number of polygons should be minimized
(to ${<}\,1\,{\rm mmag}$).

In order to obtain 1 light curve point, we thus proceed as follows.
At the stellar surface, we compute
the monochromatic intensity:
\begin{equation}
B_\lambda = {2hc^2\over\lambda^5} {1\over\exp\left({hc/(\lambda kT)}\right)-1}\,,
\end{equation}
the monochromatic flux:
\begin{equation}
\Phi_\lambda = \pi B_\lambda\,,
\end{equation}
the monochromatic power:
\begin{equation}
P_\lambda = 4\pi R_{\rm S}^2\Phi_\lambda\,,
\end{equation}
and the pass-band power:
\begin{equation}
P_V = \Delta_{\rm eff} P_\lambda\,,
\end{equation}
where $\Delta_{\rm eff}$ denotes the effective passband.

At the asteroid surface, we compute
the incoming monochromatic flux:
\begin{equation}
\Phi_\lambda = {P_\lambda\over 4\pi d_1^2}\,,
\end{equation}
where $d_1$ denotes the Sun--asteroid distance;
the pass-band flux:
\begin{equation}
\Phi_V = \Delta_{\rm eff} \Phi_\lambda\,,
\end{equation}
the reflectance for the given spectral slope~$\gamma$:
\begin{equation}
{\cal R} = 1 + \gamma(\lambda_{\rm eff}/(1\,\mu{\rm m}) - 0.55)\,,
\end{equation}
and the Lambert law factor:
\begin{equation}
f_{\rm L} = {\cal R}{A_{\rm w}\over 4\pi}\,,
\end{equation}
where $A_{\rm w}$ denotes the single-particle albedo (at $0{,}55\,\mu{\rm m}$).

At the observer location, we evaluate
the solid angle:
\begin{equation}
\omega = {1\over d_2^2}\,,
\end{equation}
where $d_2$ denotes the asteroid--observer distance;
and the pass-band calibration flux:
\begin{equation}
\Phi_{V,{\rm cal}} = \Delta_{\rm eff}\Phi_{\lambda,{\rm cal}}\,.
\end{equation}

Originally, the shape model is composed of triangular faces.
A scaling of nodes,
axis rotations,
pole orientations,
and relative positions of both (22) and Linus
are subsequently computed.

A conversion to $i$ sets of polygons $s_i$ is then performed;
this is important because a clipping of one triangle by another triangle
is one or more polygons.
Every set contains $j$ polygons $p_{i,j}$.
Every polygon contains $k$ points $\vec p_{i,j,k}$.
Yet, each set $s_i$ is always located in the same plane,
because we retain the original geometry.

The geometry is described by
the normals $\hat n$,
the centres $\vec c$, and
the directional cosines $\mu_{\rm i}$, $\mu_{\rm e}$.
The non-illuminated and non-visible polygons won't be computed
($\mu_{\rm i}\le 0 \land \mu_{\rm e}\le 0$).

The 1st transformation is determined by the asteroid$\to$Sun unit vector $\vec s$,
which determines the new basis:
\begin{equation}
\hat w = \vec s\,,\quad
\hat u = (-\sin l, \cos l, 0)\,,\quad
\hat v = -\hat u\times\hat w\,,
\end{equation}
and the respective coordinates:
\begin{equation}
(u, v, w)_{i,j,k} = (\hat u\cdot\vec p_{i,j,k}, \hat v\cdot\vec p_{i,j,k}, \hat w\cdot\vec p_{i,j,k})\,.\label{uvw}
\end{equation}
To optimize the computation, we perform bounding-box tests.
Only if polygons are in proximity,
we compute the 2-dimensional shadowing (clipping)
with the 3-dimensional back-projection as:
\begin{equation}
z = {d - ax - by\over c}\,,\label{z}
\end{equation}
where $(a, b, c) \equiv \hat n$,
and $d = \hat n\cdot\vec c$.

The 2nd transformation is determined similarly
by the asteroid$\to$observer unit vector $\vec o$,
with the same Eq.~(\ref{uvw}).
We then compute the visibility (clipping),
and the back-projection,
with the same Eq.~(\ref{z}).
The surface area of the resulting polygons is computed as:
\begin{equation}
\vec S_j = \sum_{k\le 2} (\vec b-\vec a)\times(\vec c-\vec a)\,,
\end{equation}
\begin{equation}
S_i = \sum_j {1\over 2} |\vec S_j|\,{\rm sgn}(\vec S_j\cdot\hat n)\,,
\end{equation}
where $\vec a$ is the 1st of the polygon points $\vec p_{i,j,1}$
and $\vec b$, $\vec c$ are the 2nd, 3rd, etc. $\vec p_{i,j,k}$, $\vec p_{i,j,k+1}$.
The sign test is necessary for small polygons $\ll$ big polygons
(or annular eclipses).

The incoming monochromatic flux (in ${\rm W}\,{\rm m}^{-2}\,{\rm m}^{-1}$) is then:
\begin{equation}
\Phi_{\rm i} = \Phi_\lambda\mu_{\rm i}\,,
\end{equation}
the monochromatic intensity (in ${\rm W}\,{\rm m}^{-2}\,{\rm sr}^{-1}\,{\rm m}^{-1}$)
is determined by the bi-directional scattering function:
\begin{equation}
I_\lambda = f(f_{\rm L}, \mu_{\rm i}, \mu_{\rm e}, \alpha) \Phi_{\rm i}\,,
\end{equation}
depending on the cosines and the phase angle~$\alpha$.
The outgoing monochromatic flux is:
\begin{equation}
\Phi_{\rm e} = I_\lambda\mu_{\rm e}\,.
\end{equation}
Finally, the integration over the surface determines
the monochromatic luminosity (in ${\rm W}\,{\rm sr}^{-1}\,{\rm m}^{-1}$):
\begin{equation}
J_\lambda = \sum_i \Phi_{{\rm e},i} S_i\,,\label{J_lambda}
\end{equation}
the pass-band flux:
\begin{equation}
\Phi_V = \Delta_{\rm eff}\omega J_\lambda\,,
\end{equation}
and the brightness (in mag):
\begin{equation}
V_0 = 0 - 2.5\log_{10}{\Phi_V\over\Phi_{V,{\rm cal}}}\,.
\end{equation}

\begin{figure}
\centering
\includegraphics[width=7cm]{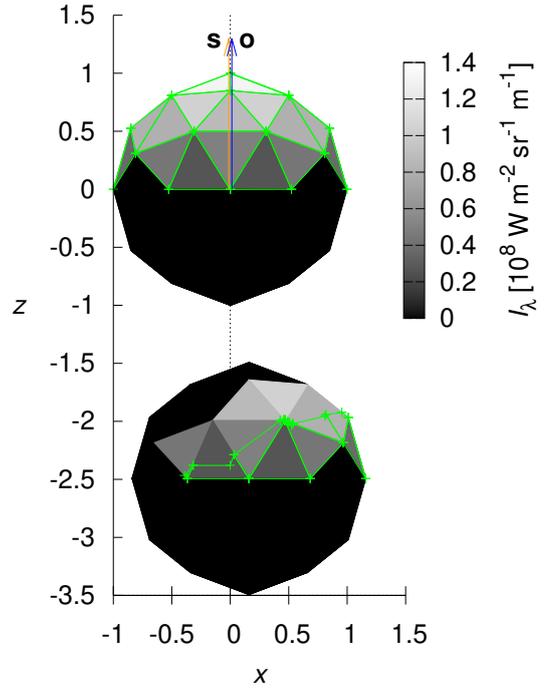}
\caption{
A 2-sphere test of the polygon light curve algorithm.
Even a very coarse discretisation, i.e., 42 nodes for each sphere, allows to compute
partial eclipses,
partial occultations, or
partial transits.
Shades of gray show
the monochromatic intensity $I_\lambda$ (in ${\rm W}\,{\rm m}^{-2}\,{\rm sr}^{-1}\,{\rm m}^{-1}$),
\color{green}green\color{black}\ lines
non-eclipsed and non-occulted polygons
used to compute the surface areas.
The \color{orange}orange\color{black}\ arrow shows the direction towards the Sun
and \color{blue}blue\color{black}\ towards the observer.
The test bodies are metre-sized,
1\,au from the Sun,
1\,au from the observer.
See also Fig.~\ref{test_polygon0_2spheres_chi2_LC2}.
}
\label{test_polygon0_2spheres_output.I_lambda.01}
\end{figure}

\begin{figure}
\centering
\includegraphics[width=8.8cm]{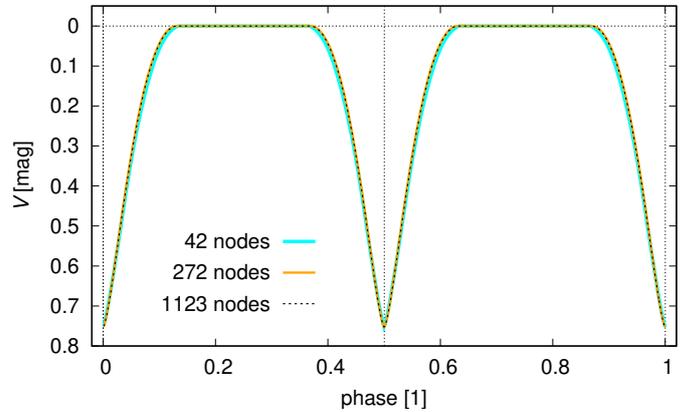}
\caption{
Light curves for a 2-sphere test,
computed for different discretisations:
42, 272, 1123 nodes.
The precision is of the order ${<}\,0.1\,{\rm mmag}$,
even for the coarse discretisation.
Tiny changes of the derivative are related to subsequently
eclipsing or occulting large triangles with different normals.
The magnitude in V band is computed for the effective wavelength
$\lambda_{\rm eff} = 545\,{\rm nm}$
and the effective passband
$\Delta_{\rm eff} = 85\,{\rm nm}$.
}
\label{test_polygon0_2spheres_chi2_LC2}
\end{figure}

\begin{figure}
\centering
\includegraphics[width=8cm]{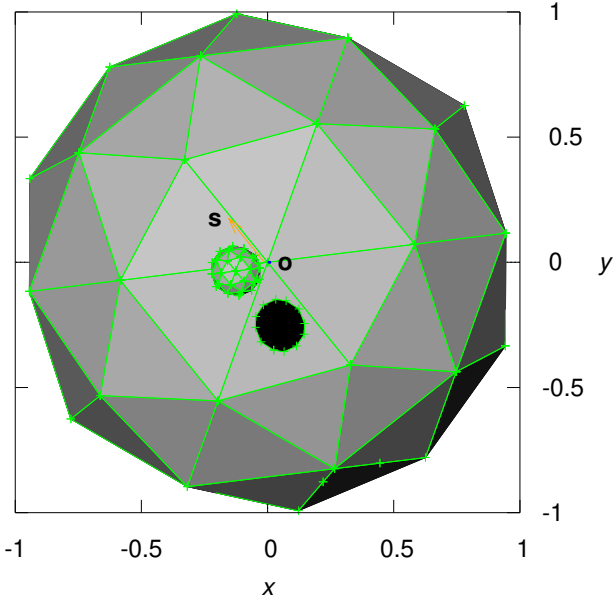}
\caption{
Similar as Fig.~\ref{test_polygon0_2spheres_output.I_lambda.01}.
A tiny-triangle test,
where one body is large and other body is small.
It demonstrates that annular eclipses, as well as
partial eclipses,
partial occultations,
partial transits,
are computed exactly.
The polygon corresponding to the shadow (black)
has a negative signed area.
}
\label{test_polygon1_tinytriangle_output.I_lambda.01}
\end{figure}

A 2-sphere test of the algorithm is demonstrated in
Fig.~\ref{test_polygon0_2spheres_output.I_lambda.01}, and
the respective light curve for different discretisations in
Fig.~\ref{test_polygon0_2spheres_chi2_LC2}.
A tiny-triangle test 
(Fig.~\ref{test_polygon1_tinytriangle_output.I_lambda.01})
shows that annular eclipses are computed exactly,
even for a~coarse discretisation.

\section{Cliptracing algorithm}\label{cliptracing}

Second, we implemented a new `cliptracing' algorithm,
which deals with the discretisation errors of synthetic images.
Traditional raytracing algorithms perform an inside-triangle test,
and use parameters of the respective triangle.
Hereinafter, we compute contributions of polygons to individual pixels exactly.

\begin{figure}
\centering
\begin{tabular}{@{}c@{\kern.3cm}c@{}}
``cliptracing'' & raytracing \\
\includegraphics[width=4.3cm]{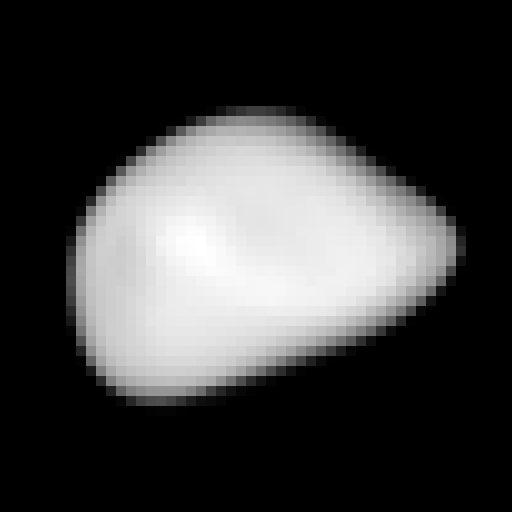} &
\includegraphics[width=4.3cm]{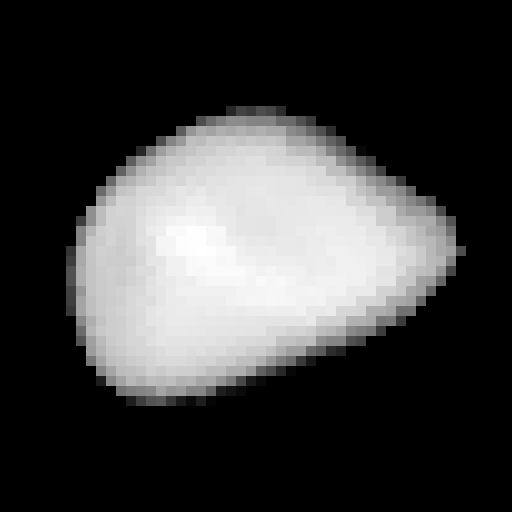} \\[0.1cm]
\end{tabular}
\caption{
A 1:1 comparison of the ``cliptracing'' (left) and the raytracing (right) algorithms.
In the former, polygons were clipped by individual pixels (analytically)
and the synthetic image of (22) is very smooth.
In the latter, a simple inside-polygon test was used for each ray,
which creates discretisation artefacts
and the synthetic image is then `noisy'.
The Lambert scattering law was used in this test.
}
\label{test_polygon13_CLIPTRACE_output_0001_syn}
\end{figure}

\begin{figure}
\centering
\begin{tabular}{@{}c@{\kern.3cm}c@{}}
\includegraphics[width=4.3cm]{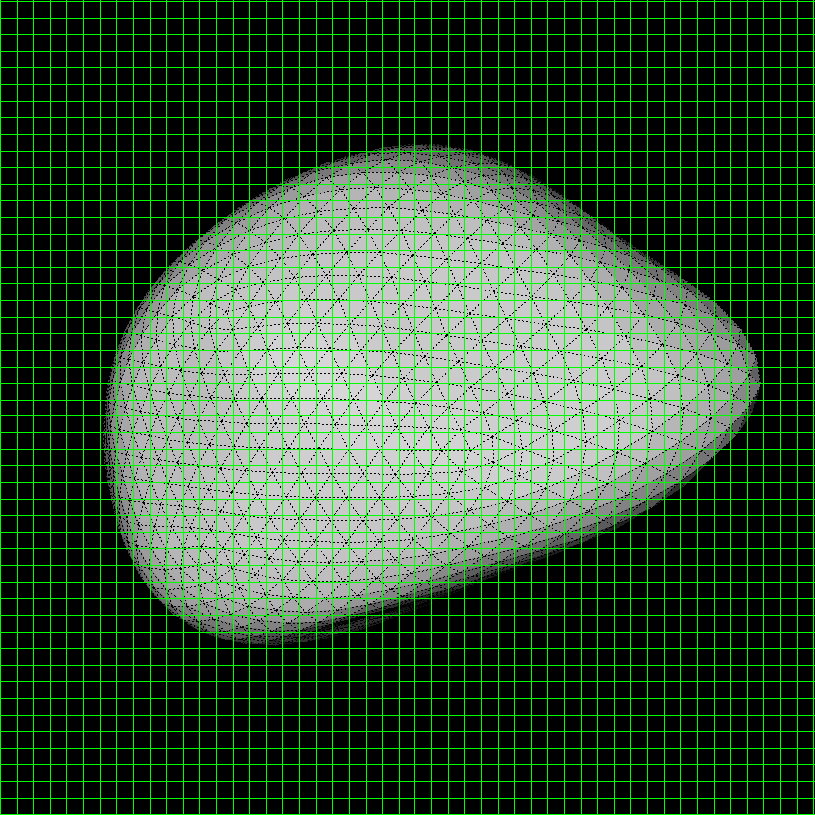} &
\includegraphics[width=4.3cm]{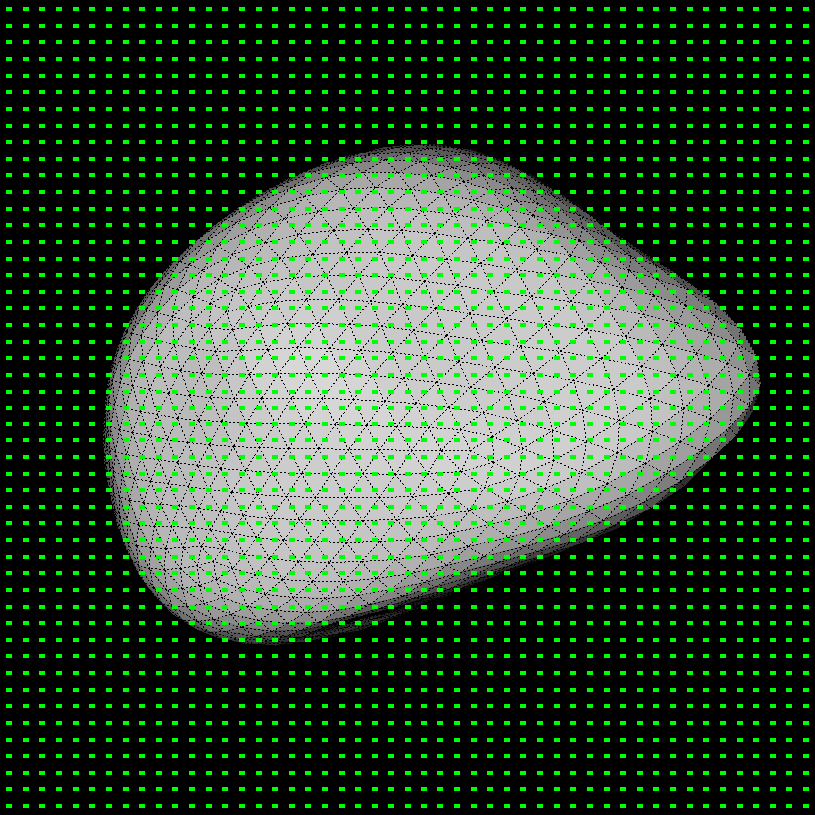} \\
\end{tabular}
\caption{
Same as Fig.~\ref{test_polygon13_CLIPTRACE_output_0001_syn},
but showing the corresponding shape composed of polygonal faces
(\color{gray}gray\color{black})
and a grid of either square pixels, or points
(\color{green}green\color{black}).
}
\label{test_polygon13_CLIPTRACE_nodes0001_GRID}
\end{figure}

We define 1~pixel as 1~polygon in the sky-plane coordinates $(u,v,w)$:
\begin{equation}
\vec p_1 = {\textstyle[(u-{1\over 2}\Delta u, v-{1\over 2}\Delta v,0), \dots, (u-{1\over 2}\Delta u, v+{1\over 2}\Delta v,0)]}\,,
\end{equation}
where
$u$, $v$ correspond to the centre of pixel,
$\Delta u$, $\Delta v$ to the size of pixel.
In a cycle over all non-shadowed $\land$ visible polygons,
we `crop' (intersect) each of them by $\vec p_1$,
and sum individual contributions to 1 pixel,
to get the monochromatic luminosity of 1 pixel:
\begin{equation}
J_\lambda = \sum_i\Phi_{{\rm e},i} {S_i'\over\mu_{{\rm e},i}}\,,
\end{equation}
where we used already projected surface area~$S_i'$,
because in our previous formalism (i.e., Eq.~(\ref{J_lambda})),
we multiply by unprojected areas $S_i$.
Everything is computed analytically,
no discretisation artefacts,
no edge artefacts,
and the outcome is a smooth synthetic image
(see Figs.~\ref{test_polygon13_CLIPTRACE_output_0001_syn},
\ref{test_polygon13_CLIPTRACE_nodes0001_GRID}).

When we compare the synthetic image with the observed one,
we have to re-center (with a sub-pixel precision).
Let us denote
$\vec c$ the photocentre of the observed AO image,
$\vec c'$ the photocentre of the synthetic AO image.
The cliptracing is therefore computed with a centre shifted to
$-\vec c+\vec c'$.

An optimisation is performed by using a number of bounding-box tests
(namely,
the observed bounding-box,
the over-all-polygons bounding-box,
the 1-pixel bounding-box,
and the individual bounding-boxes of polygons).

Another problem we have to deal with at this level of precision,
is a correlation of `everything' with the shape.
Especially the timings of events depend on details of the shape,
henceforth we created a version of our modelling tool focused
on the fitting of shape (``{\tt Xitaushp}'').

The parameters are radius vectors of the `control' shape,
which is processed by a sub-division algorithm
\citep{Kobbelt_2000,Viikinkoski_2015A&A...576A...8V},
with 1 up to 4 levels.
The input orbit is read from the previous output,
for simplicity.

To constrain the shape, we use a modified $\chi^2$ metrics:
\begin{equation}
\chi^2 = w_{\rm lc}\chi^2_{\rm lc} + w_{\rm ao}\chi^2_{\rm ao} + w_{\rm ao2}\chi^2_{\rm ao2}\,,
\end{equation}
where the individual contributions (and weights) correspond to
the light curves (LC), silhouettes (AO), and synthetic images (AO2).
If not stated otherwise, we use unit weights.
The computation of silhouettes was already described in \cite{Broz_2021A&A...653A..56B}.
It was improved by a multi-point interpolation,
which is smooth even for low resolution,
even for deconvolution artefacts,
which is occasionally present as a drop of signal at the edge (`stair case').

The synthetic image is convolved with the point-spread function (PSF).
We use the Moffat PSF:
\begin{equation}
{\rm PSF}(u,v) = {\beta-1\over\pi\alpha^2} \left(1 + {u^2+v^2\over\alpha^2}\right)^{-\beta}\,,
\end{equation}
with free parameters $\alpha$, $\beta$.
Alternatively, an observed stellar PSF can be input;
it is rather complex,
with the Strehl ratio about 0.1,
a diffraction pattern,
a ring,
a cross,
remaining AO artefacts.
However, for deconvolved images we would need a 'residual' PSF instead.

The respective $\chi^2$ contribution for synthetics images is computed
as a sum over pixels:
\begin{equation}
\chi^2_{\rm ao2} = \sum_{u,v} {(J_\lambda'-J_\lambda)^2\over\sigma^2} {\cal H}(J_\lambda-J_{\rm min}) {\cal H}(J_\lambda'-J_{\rm min})\,,
\end{equation}
where $J_\lambda(u,v)$ denotes the observed monochromatic luminosity,
$J_\lambda'$ synthetic,
${\cal H}(x)$ the Heaviside step function.
The Poisson uncertainty is computed as
$\sigma^2 = {\rm max}(J_\lambda', J_\lambda)$
for each pixel,
because sometimes we have to compare to darkness ($J_\lambda = 0$).
The minimum luminosity is computed as
$J_{\rm min} = f\,{\rm max}(J_\lambda)$
for all pixels,
because background is rather extended and uneven;
the factor~$f$ (threshold) often corresponds to the silhouettes.

\section{Stellar occultation algorithm}\label{occultation}

Third, we implemented a new occultation algorithm in {\tt Xitau}.
It is used to check the astrometric positions, timings of occultations
and precision of the ephemerides.
At the beginning, we use a sphere-intersection test
to speed-up the computation.
We apply a standard TDB to UT1 conversion
\citep{SOFA_2014ascl.soft03026I},
precession \citep{Lieske_1977A&A....58....1L}
nutation \citep{Wahr_1981GeoJ...64..705W,Wolf_1992aspr.book.....W},
an equatorial-of-J2000 to equatorial-of-date transformation,
a proper motion of the respective star (from the Gaia DR3),
and an ellipsoid-intersection test:
\begin{equation}
\vec A + x\vec B = \vec e\,,
\end{equation}
\begin{equation}
\left({e_1\over a}\right)^2 + \left({e_2\over b}\right)^2 + \left({e_3\over c}\right)^2 = 1\,,
\end{equation}
namely for the WGS-84 ellipsoid
($a = b = 6.378173\cdot 10^6\,{\rm m}$, $c = 6.3567523142\cdot 10^6\,{\rm m}$),
where
$\vec A$~denotes the Earth$\to$asteroid vector,
$\vec B$~star$\to$asteroid (normalized),
$\vec e$~the intersection point on the ellipsoid;
$x$~is a nuisance parameter.
The equation is quadratic in $x$.
At the end, we apply UT1 to GST conversion,
the Earth rotation, and
a~transformation to the geodetic coordinates:
\begin{equation}
N = {a^2\over\sqrt{(a\cos\phi)^2 + (b\sin\phi)^2}}\,,
\end{equation}
\begin{equation}
\vec e = \pmatrix{
(N+h)\cos\lambda\cos\phi \cr
(N+h)\cos\lambda\cos\phi \cr
(Nb^2/a^2+h)\sin\phi \cr
}\,,
\end{equation}
with an iterative procedure for the inverse.

\begin{figure}
\centering
\includegraphics[width=9cm]{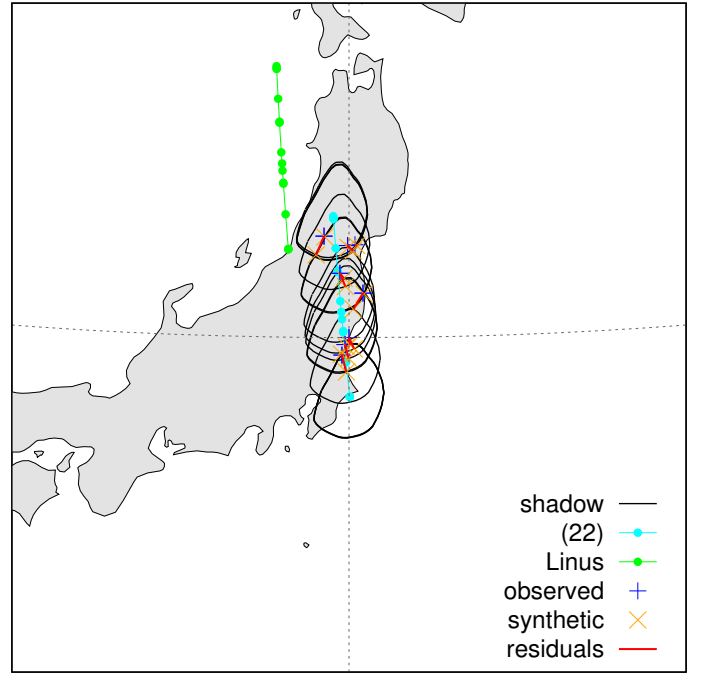}
\caption{
Nov 7th 2006 stellar occultation of (22) Kalliope and Linus,
computed for the model with $\chi^2 = 6695$ (from Tab.~\ref{tab2}).
The trajectories of (22) and Linus are plotted (\color{cyan}cyan\color{black}, \color{green}green\color{black}),
together with several shadows of (22) computed for the observed timings (black),
observers' locations (\color{blue}blue\color{black}),
synthetic locations of the nearest shadow points (\color{orange}yellow\color{black}), and
residuals (\color{red}red\color{black}).
The timings were taken from the occultation database
\citep{Herald_2019pdss.data....3H,Herald_2020MNRAS.499.4570H}.
Absolute time measurements were used to check the ephemeris accuracy.
The gnomonic projection was used;
showing Japan, Honsh$\bar{\rm u}$.
}
\label{22_test48_update__6695_chi2_OCC_xy}
\end{figure}

An example for (22)~Kalliope is shown in Fig.~\ref{22_test48_update__6695_chi2_OCC_xy}.
The algorithm was verified against selected events from the Occult software
\citep{Herald_2020MNRAS.499.4570H},
e.g., the astrometric position of (216)~Kleopatra, on Mar 12th 2015
(see Fig.~\ref{216_test79_20150312_occultation3}).

\section{Occultation, transit and eclipse events}\label{results}

Before we proceed with fitting, it is useful to summarize our dynamical model:
it uses the Bulirsch--Stoer numerical integrator,
adaptive time step,
which allows us to compute
non-keplerian orbits,
multipoles up to the order $\ell = 2$,
internal tides, or
external tides by the Sun
\citep{Broz_2017ApJS..230...19B,Broz_2021A&A...653A..56B,Broz_2022A&A...657A..76B}.
In the nominal model, we assume the tidal time lag
$\Delta t_1 \simeq 40\,{\rm s}$ inferred for (216) Kleopatra.

Modifications were described in detail in
Sects.~\ref{polygon}, \ref{cliptracing}, \ref{occultation}.
Explanation of all parameters is included in Tab.~\ref{tab2}.
In particular, we added a few free parameters, namely
$C_{20,1}$,
$A_{\rm w1}$,
$A_{\rm w2}$,
$B_0$,
$h$,
$g$,
$\bar\theta$,
which allowed us to fit the astrometry, light curves, or scattering parameters.
Alternatively, we added offsets in $\hat x$, $\hat y$, $\hat z$ directions,
due to a possible rotation about different axis.

We use several types of observations to constrain the model:
\begin{equation}
\chi^2 = w_{\rm sky}\chi^2_{\rm sky} + w_{\rm ao}\chi^2_{\rm ao} + w_{\rm lc}\chi^2_{\rm lc} + w_{\rm occ}\chi^2_{\rm occ}\,,
\end{equation}
where individual contributions correspond to
astrometry (the so-called SKY dataset),
silhouettes (AO),
light curves (LC), or
occultations (OCC);
all of them with corresponding weights.
Previously, we used $w_{\rm ao} = 0.003$ so that AO contributes comparably as SKY.
Of course, every model requires reasonable initial conditions;
therefore we used the best fit from \citet{Ferrais_2022A&A...662A..71F},
with the osculating elements adjusted to match slightly simplified
dynamics ($\ell = 2$).
We verified that omitting high-order terms does not substantially shift
the values of low-order terms, in particular, of the oblateness $C_{20}$.


\subsection{Long-arc scattering model}\label{longarc_scattering}

\begin{figure}
\centering
\includegraphics[width=9cm]{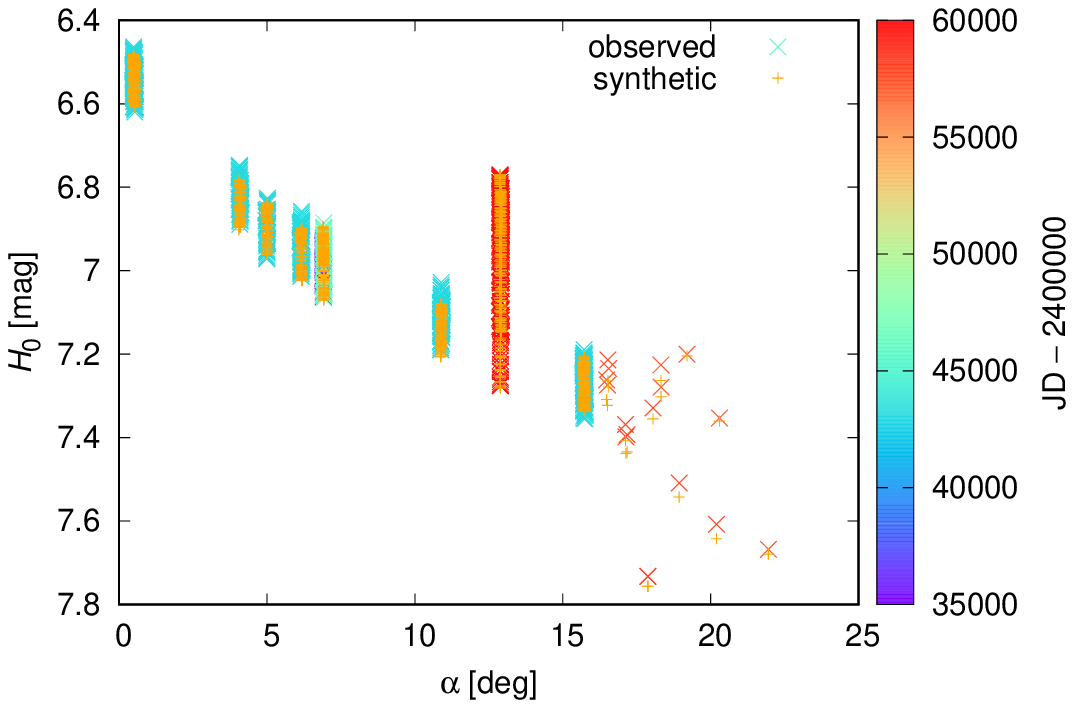}
\caption{
Phase curve of the (22) Kalliope and Linus system.
The reduced brightness $H_0$ vs. the phase angle $\alpha$ is plotted.
Calibrated photometry from
\cite{Gehrels_1962ApJ...135..906G},
\cite{Scaltriti_1978Icar...34...93S},
\cite{Surdej_1986A&A...170..167S},
this work,
and the Gaia DR3
was used.
The observed curve ($\times$) is plotted in colour (according to the Julian date),
the synthetic ($+$) in yellow.
The range of $\alpha$ is from $2$ to $22^\circ$.
The unreduced $\chi_{\rm lc}^2 = 45995$, $n_{\rm lc} = 1892$
with systematics up to 0.05\,mag.
The `scatter' of points is mostly due to the light curve,
which is fitted by our model.
Our measurements with $\alpha \sim 13^\circ$,
taken in the Cousins R band,
are in agreement with our model (0.001\,mag).
}
\label{22_test45_absolute6__46494_lightcurve2}
\end{figure}

First, we focused on the phase curve, which is controlled by
6 free parameters $A_{\rm w1}$, $A_{\rm w2}$, $B_0$, $h$, $g$, $\bar\theta$.
We used the calibrated photometry from
\cite{Gehrels_1962ApJ...135..906G},
\cite{Scaltriti_1978Icar...34...93S},
\cite{Surdej_1986A&A...170..167S},
this work, and
the Gaia DR3 data.
The zero points were fixed,
however, we expect calibration systematics up to 0.05\,mag.
A shape model of (22) Kalliope is necessary,
because the lightcurve amplitude can reach 0.6\,mag,
depending on geometry.
The simplex and subplex
(i.e., simplex on subspaces; \citealt{Rowan_1990})
algorithms were used, with several restarts.
The resulting phase curve is shown in
Fig.~\ref{22_test45_absolute6__46494_lightcurve2}.

The model is sensitive mostly to $A_{\rm w1}$, $B_0$, $h$ scattering parameters.
The unreduced $\chi^2_{\rm lc} = 45955$ is too large
compared to the number of observations $n_{\rm lc} = 1892$
due to remaining calibration systematics.
If 0.05\,mag uncertainties are used instead,
the $\chi^2$ decreases down to $n$.
Nevertheless, the best-fit values,
$A_{\rm w1} = 0.419$,
$B_0 = 1.733$,
$h = 0.0295$,
seem to be reasonable.
A correlation exists between $g$, $A_{\rm w1}$, $B_0$,
because small $g$ can be compensated by large $A_{\rm w1}$, $B_0$.
Hence, the overall uncertainties are increased to
0.10, 0.1, 0.01, respectively.
If $g$, $\bar\theta$ are also free,
their values tend to converge towards
$g \simeq 0$,
$\bar\theta \simeq 0^\circ$,
which do not seem to be common (cf.~\citealt{Li_2015aste.book..129L}).
In order to fit both the phase curve and the light curve
with the given shape of (22),
we shall use these values,
until we modify the shape (Sec.~\ref{shape}).
Their uncertainties are of the order of
0.10, $10^\circ$, respectively.
Detailed light curve shape is also sensitive to $g$, $\bar\theta$.


\subsection{Short-arc scattering model}\label{shortarc_scattering}

\begin{figure}
\centering
\includegraphics[width=8.0cm]{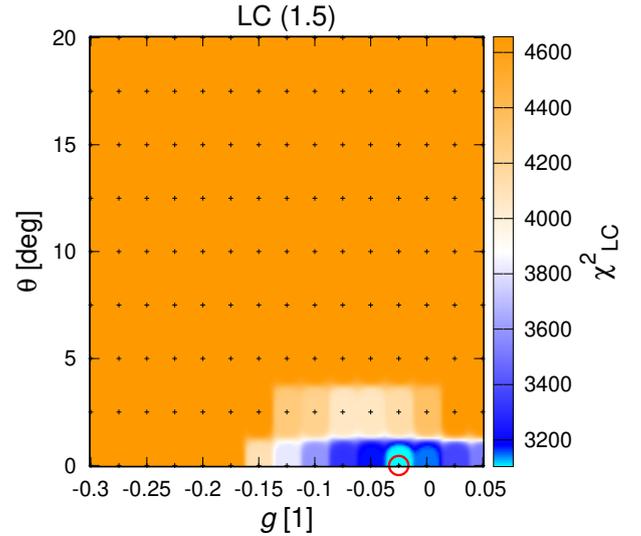}
\caption{
The asymmetry factor $g$ vs. the roughness $\bar\theta$
scattering parameters.
The corresponding $\chi^2_{\rm lc}$ values for 2 light curves are plotted as colours:
\color{cyan}cyan\color{black}\ best fits,
\color{blue}blue\color{black}\ good fits (1.2 times the best-fit $\chi^2$),
\color{orange}orange\color{black}\ poor fits (1.5).
Models were converged for 135 combinations of the fixed parameters;
other parameters were free.
The best-fit unreduced $\chi_{\rm lc}^2 = 3103$, $n_{\rm lc} = 909$ (red circle)
corresponds to
$g = -0.025$,
$\bar\theta = 0^\circ$.
Other scattering parameters were kept fixed.
}
\label{22_fitting16_MINGHAPKE__3561_ming_bartheta_LC}
\end{figure}

We fitted details on 2 light curves from SPECULOOS-Artemis (datasets 3, 4),
because we have to use calibrated photometry and avoid any zero-point
offsets between datasets. (Again, zero points were fixed.)
We computed a systematic grid for $g$, $\bar\theta$ parameters,
which were kept fixed,
while $R_1$, $R_2$, $P_{\rm rot1}$, $A_{\rm w1}$, $A_{\rm w2}$ parameters
were free.
According to Fig.~\ref{22_fitting16_MINGHAPKE__3561_ming_bartheta_LC},
the fit is still not perfect
($\chi^2_{\rm lc} = 3103$, $n_{\rm lc} = 909$)
partly because everything is interrelated ---
albedo, scattering, shape, pole, Linus, orbit, occultations, transits, eclipses, etc.
It is possible to find solutions for
$g = -0.10$ up to $0.05$, and
$\bar\theta \simeq 0^\circ$.
Small values of $g$ seem to be excluded,
because the albedo $A_{\rm w2}$ of Linus is pushed to unrealistic low values.
Large values of $\bar\theta$ seem to be excluded,
because the light curve amplitude is incorrect
(at least for the given shape).
Given the preference for negative~$g$
\citep{Spjuth_2009PhDT.......588S},
we prefer solutions close to
$g = -0.025$,
$\bar\theta \simeq 0^\circ$.
Alternatively,
albedo variegation, or
roughness variegation
may be present on the surface.



\subsection{Short-arc, astrometric + photometric model}

\begin{figure}
\centering
\includegraphics[width=9cm]{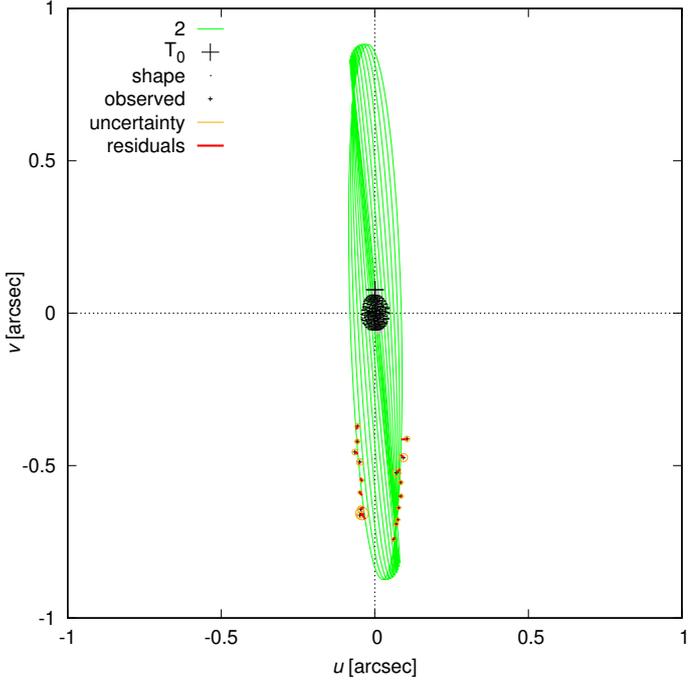}
\caption{
Orbit of Linus in the $(u,v)$ plane,
derived from the short-arc, astrometric + photometric model.
It fits the PISCO dataset around 2459579,
i.e., close to the mutual occultation events,
when the orbit is seen from the edge.
The synthetic orbit of Linus (i.e., body 2) is plotted in \color{green}green\color{black},
the observed astrometry in \color{orange}yellow\color{black},
the residuals in \color{red}red\color{black},
the shape of (22) in black.
The viewing geometry is changing in the course of time;
otherwise the orbit is elliptical.
The position at the reference epoch $T_0$ is marked by the cross.
The contribution to $\chi^2$ is $\chi^2_{\rm sky} = 27$, $n_{\rm sky} = 36$.
}
\label{22_test25_shortarc__6462_chi2_SKY_uv}
\end{figure}

\begin{figure}
\centering
\includegraphics[width=9cm]{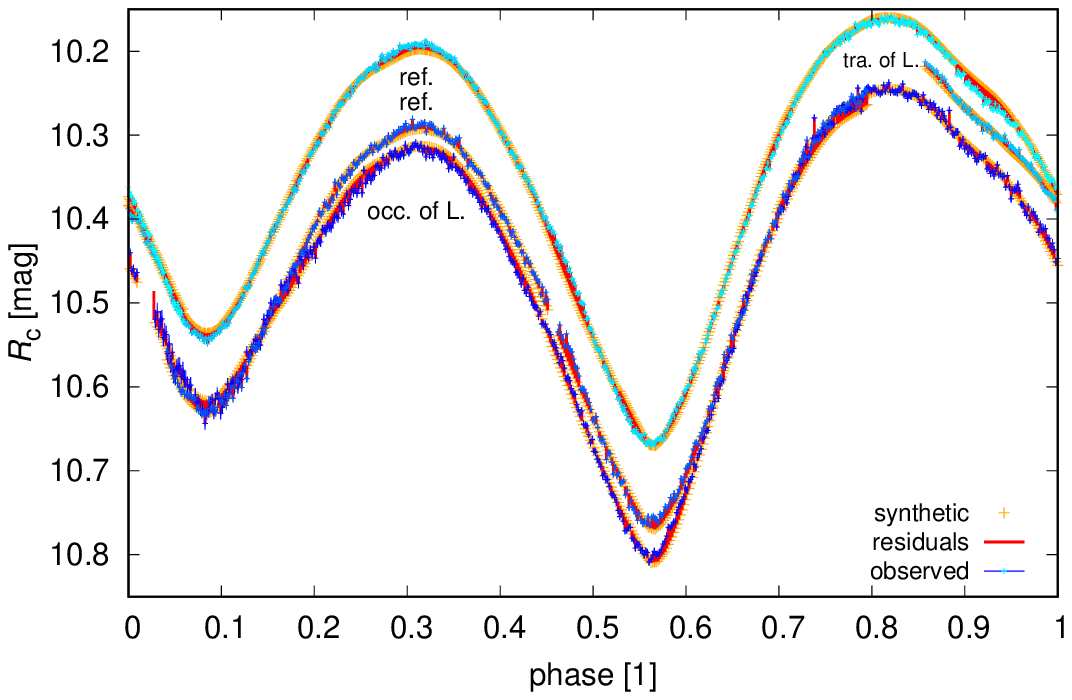}
\includegraphics[width=9cm]{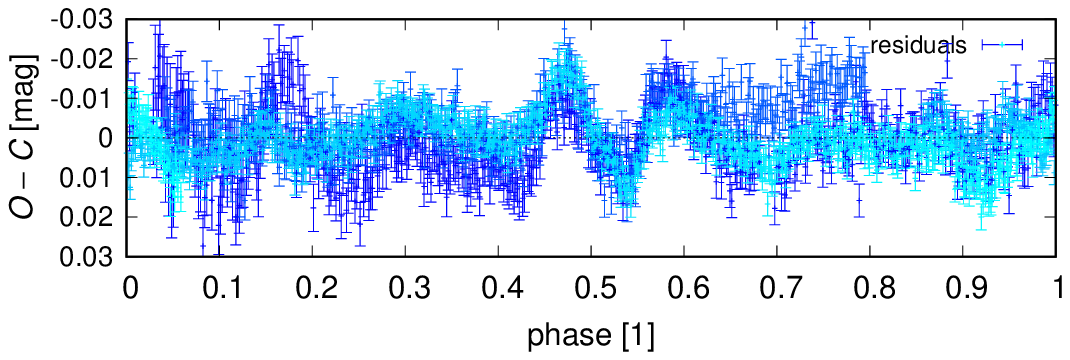}
\caption{
Phased light curves for the short-arc, astrometric + photometric model (top).
It shows the 1st occultation of Linus
and the 2nd transit of Linus,
together with the reference light curves.
The observed light curve is plotted as \color{blue}blue\color{black}\ (with error bars),
the synthetic as \color{orange}yellow\color{black},
the residuals as \color{red}red\color{black}.
The shades of blue correspond to the Julian date.
The drop in brightness is up to 0.05\,mag.
Both the amplitude and duration of the events are in agreement.
For comparison, we also plot the difference $O-C$ (bottom).
Remaining systematics occur on all light curves;
so they must be related to the shape, not the occultations.
The contribution to $\chi^2$ is $\chi^2_{\rm lc} = 6296$, $n_{\rm lc} = 1829$.
}
\label{22_test25_shortarc__6462_chi2_LC2_PHASE}
\end{figure}

\begin{figure}
\centering
\includegraphics[width=8.5cm]{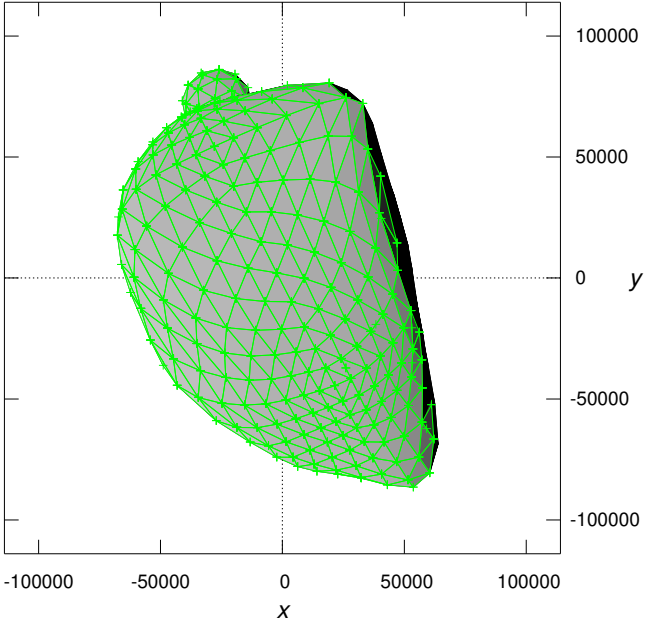}
\caption{
An example of geometry for the mutual occultation of Linus by (22) Kalliope,
namely the event 2459546.
The monochromatic intensity $I_\lambda$ (in ${\rm W}\,{\rm m}^{-2}\,{\rm sr}^{-1}\,{\rm m}^{-1}$)
is shown as shades of gray.
The ADAM shape model with 800 faces was used for (22),
and a sphere with 80 faces for Linus.
It is sufficient,
because partial occultations of faces were computed by the polygonal light curve algorithm.
}
\label{22_test25_shortarc__6462_output.I_lambda.20}
\end{figure}

As the next step, we fitted 4 light curves from SPECULOOS-Artemis and TRAPPIST-South
(datasets 1, 2, 3, 4),
together with the PISCO astrometric dataset,
which was acquired very close to the occultation and transit events.
Also the silhouettes were used to prevent incompatible pole orientations,
which influence the events.
Analytical zero points were computed, compensating for remaining offsets between the respective light curves.
The best-fit total weighted unreduced $\chi^2 = 6462$,
with the individual contributions
$\chi^2_{\rm sky} = 27$,
$\chi^2_{\rm lc} = 6296$, and
$\chi^2_{\rm ao} = 45583$,
where the respective numbers of observations
$n_{\rm sky} = 36$,
$n_{\rm lc} = 1829$,
$n_{\rm ao} = 12960$.
The fit exhibits no systematics in astrometry;
see Fig.~\ref{22_test25_shortarc__6462_chi2_SKY_uv}.
Minor systematics are present in the light curves,
(Fig.~\ref{22_test25_shortarc__6462_chi2_LC2_PHASE}),
nevertheless, the amplitude as well as the duration of the occultations and transit events
is matched almost perfectly.
An example of geometry is shown in Fig.~\ref{22_test25_shortarc__6462_output.I_lambda.20}.

We checked that a mirror solution (occultation$\,\leftrightarrow\,$transit)
is not possible;
the variable geometry allows to distinguish these solutions.


\subsection{Long-arc, astrometric model}\label{longarc_astrometric}

In order to constrain the dynamical parameters,
astrometric measurements from 2452151 to 2459580 were used,
as well as the silhouettes to prevent incompatible pole orientations.
In this case, the free parameters were:
$m_{\rm sum}$,
$P_1$,
$\log e_1$,
$i_1$,
$\Omega_1$,
$\varpi_1$,
$\lambda_1$,
$l_{\rm pole1}$,
$b_{\rm pole1}$,
$\phi_{01}$,
while the fixed parameters:
$q_1$,
$C_{20,1}$,
$\Delta t_1$.
Actually, we computed an extended grid for the latter two parameters,
in the range of
$-0.22$ to $-0.08$,
$0$ to $60\,{\rm s}$,
respectively.
Apart from $C_{20,1}$, we included other multipoles up to $\ell = 2$,
which were computed for a homogeneous structure:

\begin{center}
\begin{tabular}{rrrr}
$C_{21}$ & $-4.365949\cdot 10^{-3}$ & $S_{21}$ & $-2.414236\cdot 10^{-3}$ \\
$C_{22}$ & $ 4.732558\cdot 10^{-2}$ & $S_{22}$ & $ 3.357381\cdot 10^{-5}$ \\
\end{tabular}
\end{center}

A very important result is that two solutions exists for the oblateness $C_{20,1}$,
either ${\simeq}\,-0.20$,
or ${\simeq}\,-0.12$
(see Fig.~\ref{22_fitting10_GRIDC20__249_C201_Deltat1_SKY}).
A homogeneous body with $C_{20,1} = -0.1586$ is excluded.
These two solutions correspond to
3~or 2~nodal precession cycles
(see, e.g., Fig.~\ref{22_test48_update__6695_orbit1}).
There is no other option (4~or 1~cycle),
because $C_{20,1}$ would be unrealistic (too high or too low).
The best-fit value is $\chi^2_{\rm sky} = 249$,
or alternatively $\chi^2_{\rm sky} = 260$.
It indicates no systematics,
possibly overestimated uncertainties,
because the number of data points is
$n_{\rm sky} = 344$
(both $\rho$, $\theta$).

We also checked a range of $\log e_1$, $i_1$ values
(Fig.~\ref{22_fitting2_GRIDECC__310_loge1_i1_SKY}).
We tested mirror solutions, retrograde solutions, shifted-by-$180^\circ$ solutions.
There is no alternative solution,
neither for the eccentricity, nor for the inclination.
The uncertainties are up to $0.5$ (in log-scale), $0.5^\circ$.



\subsection{Long-arc, astrometric + photometric model}\label{longarc_astrometric_photometric}

In order to constrain the physical parameters,
4 light curves (datasets 1, 2, 3, 4),
all astrometric measurements, and
all silhouettes
were used.
Apart from dynamical parameters,
additional free parameters
$R_1$, $R_2$, $P_{\rm rot1}$, $A_{\rm w2}$
can be now constrained by mutual occultation and transit events.
We fixed the parameters:
$A_{w1}$,
$B_0$,
$h$,
$g$,
$\bar\theta$,
otherwise the model would not match the calibrated photometry (Sec.~\ref{longarc_scattering}).
Again a grid of
$C_{20,1}$,
$\Delta t_1$
was computed
(Fig.~\ref{22_fitting12_GRIDLC__7095_C201_Deltat1_ALL}).

The best-fit $\chi^2 = 7095$,
with the individual contributions
$\chi^2_{\rm sky} = 337$,
$\chi^2_{\rm lc} = 6600$, and
$\chi^2_{\rm ao} = 51856$.
All of them are slightly worse,
most likely due to a combination of more observational datasets,
but it is an acceptable compromise.

Regarding the oblateness,
$C_{20,1} \simeq -0.12$ seems to be a bit more compatible with the light curves,
but we still cannot exclude the $-0.22$ solution.
The tidal time lag is not well constrained.
The volume-equivalent diameters of (22),
$D_1 = 151.0\,{\rm km}$,
is still compatible with the ADAM or MPCD shapes
\citep{Ferrais_2022A&A...662A..71F};
the uncertainty is of the order of $1\,{\rm km}$.

\begin{figure}
\centering
\includegraphics[width=8cm]{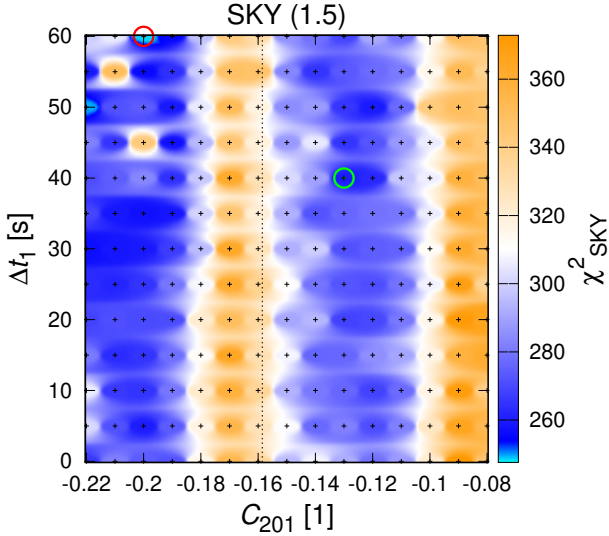}
\caption{
The quadrupole moment $C_{20,1}$
vs. the tidal time lag $\Delta t_1$
of the central body.
The corresponding $\chi^2_{\rm sky}$ values
are plotted as colours (cyan, blue, white, orange)
and as numbers (gray).
SKY and AO datasets were used.
Models were converged for
195 combinations of the fixed parameters;
all other parameters were free.
For each combination, 1000 iterations were computed, i.e., 195000 models in total.
Homogeneous body with $C_{20,1} = -0.1586$ is excluded.
Preferred solutions are either ${\simeq}-0.20$, or ${\simeq}-0.12$, indicated by
\color{red}red\color{black}\ and
\color{green}green\color{black}\ circles.
}
\label{22_fitting10_GRIDC20__249_C201_Deltat1_SKY}
\end{figure}

\begin{figure}
\centering
\includegraphics[width=8.9cm]{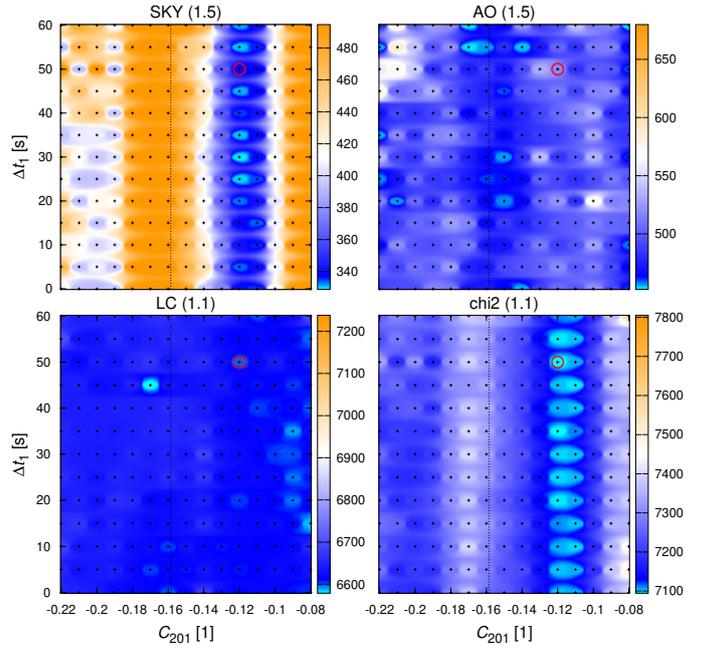}
\caption{
Similar as Fig.~\ref{22_fitting10_GRIDC20__249_C201_Deltat1_SKY}
for three datasets SKY, AO, LC, and the total~$\chi^2$.
A subset of 4 light curves was used.
The LC dataset allows to constrain the respective parameters ($C_{20,1}$, $\Delta t_1$) even better,
because 1 occultation and 1 transit
must have a specific geometry.
The AO contribution is mostly blue,
but it does not mean that this dataset is unimportant.
Actually, it excludes a lot of models with incompatible poles.
}
\label{22_fitting12_GRIDLC__7095_C201_Deltat1_ALL}
\end{figure}

\begin{figure}
\centering
\includegraphics[width=8.9cm]{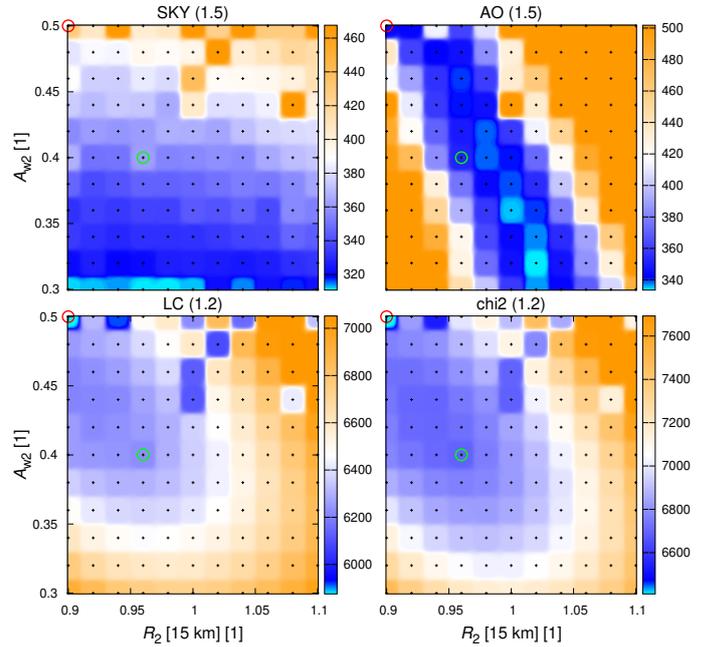}
\caption{
Similar as Fig.~\ref{22_fitting10_GRIDC20__249_C201_Deltat1_SKY}
for three datasets SKY, AO, LC.
The radius $R_2$ vs. the single-particle albedo $A_{\rm w2}$ of Linus is plotted.
Models were converged for 121 combinations of the fixed parameters;
all other parameters were free.
The preferred solution has
$\chi^2_{\rm sky} = 362$, $n_{\rm sky} = 344$,
indicated by \color{green}green\color{black}\ circle.
The radius is given in the unit of nominal radius (15\,km);
the best-fit diameter is then $27.6\,{\rm km}$.
}
\label{22_fitting15_R2A2__6695_R2_albedo2_ALL}
\end{figure}

A specific grid was computed for $R_2$, $A_{\rm w2}$ of Linus
(Fig.~\ref{22_fitting15_R2A2__6695_R2_albedo2_ALL}).
It further improved $\chi^2 = 6695$.
In this particular model,
$D_2 = 28.8\,{\rm km}$,
with a similar uncertainty.
It is compatible with the stellar occultation
observed on Nov 7th 2006
\citep{Descamps_2008Icar..196..578D}.
All parameters of this model are presented in Tab.~\ref{tab2}.
The global uncertainties of parameters were estimated
from a series of alternative admissible solutions
(cf. models above;
Figs.~\ref{22_fitting10_GRIDC20__249_C201_Deltat1_SKY},
\ref{22_fitting12_GRIDLC__7095_C201_Deltat1_ALL},
\ref{22_fitting15_R2A2__6695_R2_albedo2_ALL}).
In this order-of-magnitude estimate, we included also a contribution
from systematic uncertainties.
We verified these results by using all available light curves (from Tab.~\ref{tab1}),
which resulted in a statistically equivalent model (cf.~Tab.~\ref{taba1}).
The Markov Chain Monte Carlo (MCMC) simulation
is presented in Fig.~\ref{22_test48_update__6695_corner_};
it demonstrates typical local uncertainties and correlations
in the surroundings of one local minimum (close to $\chi^2 = 6695$).



\begin{figure}
\centering
\includegraphics[width=9cm]{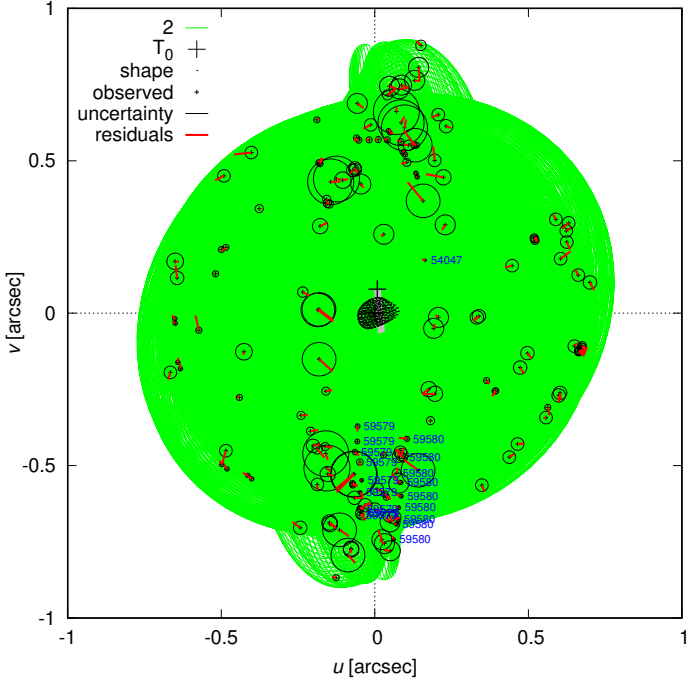}
\caption{
Same as Fig.~\ref{22_test25_shortarc__6462_chi2_SKY_uv},
for the long-arc, astrometric + photometric model.
It was constrained by the astrometry over the time span 2001--2022.
The most important measurements are indicated by \color{blue}blue\color{black}\ labels
($\hbox{Julian date} - 2400000$).
The contribution to $\chi^2$ is $\chi^2_{\rm sky} = 364$, $n_{\rm sky} = 344$.
}
\label{22_test48_update__6695_chi2_SKY_uv}
\end{figure}

The Linus' orbit, together with available astrometric measurements, is shown in
Fig.~\ref{22_test48_update__6695_chi2_SKY_uv}.
It was checked by the stellar occultation computation
(Fig.~\ref{22_test48_update__6695_chi2_OCC_xy}).
The temporal evolution of osculating elements is demonstrated in
Fig.~\ref{22_test48_update__6695_orbit1}.

In most solutions, Linus seems to be darker compared to (22),
the single-scattering albedo is
$A_{\rm w2} = 0.400$,
$A_{\rm w1} = 0.438$,
respectively,
with the local uncertainties of the order of $0.02$.
They are naturally correlated.
The darkness is also apparent on those deconvolved AO images,
which capture both (22) and Linus at the same time.
The appearance is only partly affected
by a difference between (more) flat vs. (more) curved surface.

Finally, we should not be misled by sparse local minima,
which are relatively deep;
they compensate some systematics on the light curve
which are however unrelated to the mutual occultation or transit events.
Yet, such solutions are in contradiction with the PISCO astrometric dataset;
see Fig.~\ref{22_test48_update__6695_chi2_SKY_uv}.
We thus prefer solutions, which fit the orbit just prior/posterior of the events.
Possibly, the shape of (22) Kalliope should be also adjusted.

\begin{table*}
\caption{Parameters, their values and uncertainties for the nominal
and high-oblateness models of the (22) Kalliope and Linus system.}
\label{tab2}
\centering
\begin{tabular}{lllll}
& nominal & high-oblateness \\
var. & val. & val. & unit & $\sigma$ \\
\hline
\vrule width 0pt height 10pt
$m_{\rm sum}$		& $3.902028\cdot 10^{-12}	$ & $3.902434\cdot 10^{-12}	$ & $M_{\rm S}$	& $0.001000\cdot 10^{-12}	$	\\
$q_1$			& $6.129\cdot 10^{-3}	 	$ & $6.129\cdot 10^{-3}	 	$ & 1		& $1.000\cdot 10^{-3}	 	$	\\
$P_1$			& $3.601774    		 	$ & $3.606096     		$ & day		& $0.000001     		 	$	\\
$\log e_1$		& $-2.195     			$ & $-2.444     		$ & 1		& $0.100     			$	\\
$i_1$			& $ 88.774     	 		$ & $ 89.130     	 	$ & deg		& $1.0       	 		$	\\
$\Omega_1$		& $373.127     	 		$ & $374.232     	 	$ & deg		& $1.0       	 		$	\\
$\varpi_1$		& $132.259     		 	$ & $129.868     		$ & deg		& $1.0     		 	$	\\
$\lambda_1$		& $359.793      		$ & $360.443      		$ & deg		& $1.0      			$	\\
$R_1$			& $0.993     			$ & $0.999     			$ & 76.5\,km	& $0.02     			$	\\
$R_2$			& $0.960     			$ & $0.955     			$ & 15\,km	& $0.02     			$	\\
$P_{\rm rot1}$		& $0.172841     		$ & $0.172841     		$ & day		& $0.000001     		$	\\
$P_{\rm rot2}$		& $3.595713^{\rm f}   		$ & $3.595713      		$ & day		& $1.0      			$	\\
$\Delta t_1$		& $50.1     		 	$ & $60.0     		 	$ & s		& $20.0     		 	$	\\
$C_{20,1}$		& $-0.1199    			$ & $-0.2000    		$ & 1		& $0.0100     			$	\\
$l_{\rm pole1}$		& $193.805     	 		$ & $194.993     	 	$ & deg		& $1.0     	 		$	\\
$b_{\rm pole1}$		& $2.515	      		$ & $1.751      		$ & deg		& $1.0      			$	\\
$\phi_{01}$		& $84.496      			$ & $86.776      		$ & deg		& $1.0      			$	\\
$A_{\rm w1}$		& $0.438     			$ & --          		  & 1		& $0.020     			$	\\
$A_{\rm w2}$		& $0.400     			$ & --          		  & 1		& $0.020     			$	\\
$B_0$			& $1.733      			$ & --           		  & 1		& $0.100      			$	\\
$h$			& $0.0295 			$ & --       			  & 1		& $0.0010 			$	\\
$g$			& $-0.0197			$ & --       			  & 1		& $0.0010			$	\\
$\bar\theta$		& $0.000      			$ & --           		  & deg		& $5.0      			$	\\
\hline
\vrule width 0pt height 10pt
$n_{\rm sky}$		& 344				& 344	\\
$n_{\rm ao}$		& 12600				& 12600	\\
$n_{\rm lc}$		& 1829				& --	\\
$n$			& 14773				& 12944 \\
$\chi^2_{\rm sky}$	& 364				& 249	\\
$\chi^2_{\rm ao}$	& 35209				& 36818	\\
$\chi^2_{\rm lc}$	& 6223				& --	\\
$\chi^2$		& 6695				& 359	\\
\hline
\end{tabular}
\tablefoot{
$m_{\rm sum}$ denotes the sum of masses,
$q_1$, mass ratio $m_2/m_1$,
$P_1$, osculating orbital period,
$\log e_1$, logarithm of eccentricity,
$i_1$, inclination with respect to the ecliptic system,
$\Omega_1$, longitude of the ascending node,
$\varpi_1$, longitude of the pericentre,
$\lambda_1$, true longitude,
$R_i$, radius of the $i$-th body
$P_{{\rm rot}i}$, rotation period,
$\Delta t_1$, tidal time lag,
$C_{20,1}$, zonal multipole coefficient,
$l_{{\rm pole}1}$, ecliptic longitude of the rotational pole,
$b_{{\rm pole}1}$, latitude of the pole,
$\phi_{01}$, rotation phase,
$A_{{\rm w}i}$, single-scattering albedo,
$B_0$, opposition effect amplitude,
$h$, opposition effect width,
$g$, asymmetry factor,
$\bar\theta$, surface roughness,
$n$, total number of observations,
$\chi^2$, total weighted unreduced $\chi^2$,
with individual contributions from the SKY, AO, LC datasets;
the respective weights
$w_{\rm sky} = w_{\rm lc} = 1$, and
$w_{\rm ao} = 0.003$
(i.e., for regularisation).
All orbital elements are osculating, for the epoch
$T_0 = 2459546.692102\,{\rm (TDB)}$.
$^{\rm f}$~denotes a fixed parameter.
}
\end{table*}


\subsection{Possible centre-of-mass offset of (22)}

The simplest adjustment is a centre-of-mass offset
due to an asymmetric internal structure.
This implies rotation about a different axis
and different extent of the central body in different directions.
Because the orbit of Linus is well constrained by astrometry,
the timings of mutual occultations or transits offer an opportunity
to measure the offset not directly (in space), but indirectly (in time).

Therefore, we added three more free parameters to our model,
the offsets in $\hat x$, $\hat y$, or $\hat z$ directions.
After testing values up to $20\,{\rm km}$,
which did affect $\chi^2$ substantially,
we conclude that it did not converge to a unique significant minimum.
We thus have to look for a more complex solution.



\subsection{Possible shape adjustment of (22)}\label{shape}

In order to have a full control,
we derived a new shape model of (22) Kalliope.
We started with a low-resolution ellipsoid,
(with the number of nodes $n = 42$),
which serves as a `control' shape.
After 3 sub-divisions,
we obtained a high-resolution shape ($n = 1082$),
which was converged and constrained by AO silhouettes
as well as light curves.
We used no additional regularisation term in our $\chi^2$ metric.
Nevertheless, we assured that $\chi^2$ is indeed sensitive to
all shape parameters.

In order to capture fine details,
we used the medium-resolution (22)-like shape ($n = 122$),
after the 1st subdivision.
After 2 more sub-divisions,
we again obtained a high-resolution shape,
which was constrained by AO images
and light curves.

In particular, we converged the following free parameters:
$R_{{\rm shp},i}$ radii of control nodes,
$R_1$,
$P_{\rm rot1}$,
$l_{\rm pole1}$,
$b_{\rm pole1}$,
$\phi_{\rm 01}$,
$A_{\rm w1}$,
$g$,
$\bar\theta$,
with several restarts of simplex or subplex.
The best-fit shape model is shown in
Fig.~\ref{22_test54_aoellipsoid__23663_chi2_AO2}.

The total signal contained in all pixels in all images is `huge',
the unreduced $\chi^2_{\rm ao2} \simeq 1.6\cdot 10^7$,
while the number of points (pixels) taken into account is $n_{\rm ao2} = 22843$.
This is at least partly due to rotation,
which changes the projected shape during an exposure in a non-trivial way,
remaining deconvolution artefacts,
present on images taken at
2458643,
2458661,
2458672,
or relatively low background level on the same set of images.


Overall, our new shape is still similar to the ADAM model
(Fig.~\ref{22_test51_ellipsoid__4195_shape}),
and it is even more so in the line-of-sights projections.
However, it is fine-tuned to the respective datasets,
with the respective contributions,
$\chi^2_{\rm ao} = 21515$ vs. $35209$,
$\chi^2_{\rm lc} = 3980$ vs. $6223$,
significantly improved.
Systematics on the light curves related to the shape were at least partly eliminated
(cf.~Fig.~\ref{22_test51_ellipsoid__4195_chi2_LC2_PHASE}).

In principle, one should use this shape and start over again
(from Sect.~\ref{longarc_scattering}, \ref{shortarc_scattering}, \dots).
While it is not beyond the scope of this paper,
we postpone such an in-depth study
--- including optimisations of `everything' together with shape ---
as a future work.




\begin{figure}
\centering
\includegraphics[width=9cm]{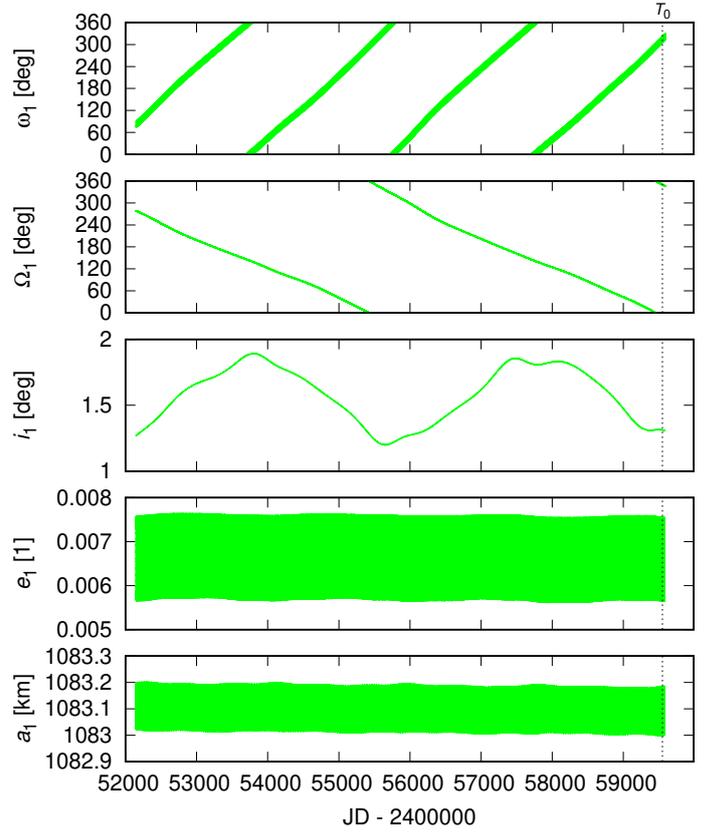}
\caption{
Evolution of osculating orbital elements of Linus
for the model with $\chi^2 = 6695$ (from Tab.~\ref{tab2}).
From bottom to top:
$a_1$~semimajor axis,
$e_1$~eccentricity,
$i_1$~inclination,
$\Omega_1$~longitude of the ascending node,
$\varpi_1$~longitude of pericentre.
The reference frame is related to the equator of (22) Kalliope;
the epoch $T_0 = 2459546.692102\,{\rm (TDB)}$.
The model includes multipoles ($\ell = 2$),
internal tides,
and external tides.
Over the time span of astrometric observations (2001--2021),
it exhibits 2 nodal precession cycles.
}
\label{22_test48_update__6695_orbit1}
\end{figure}

\begin{figure}
\centering
\includegraphics[width=9cm]{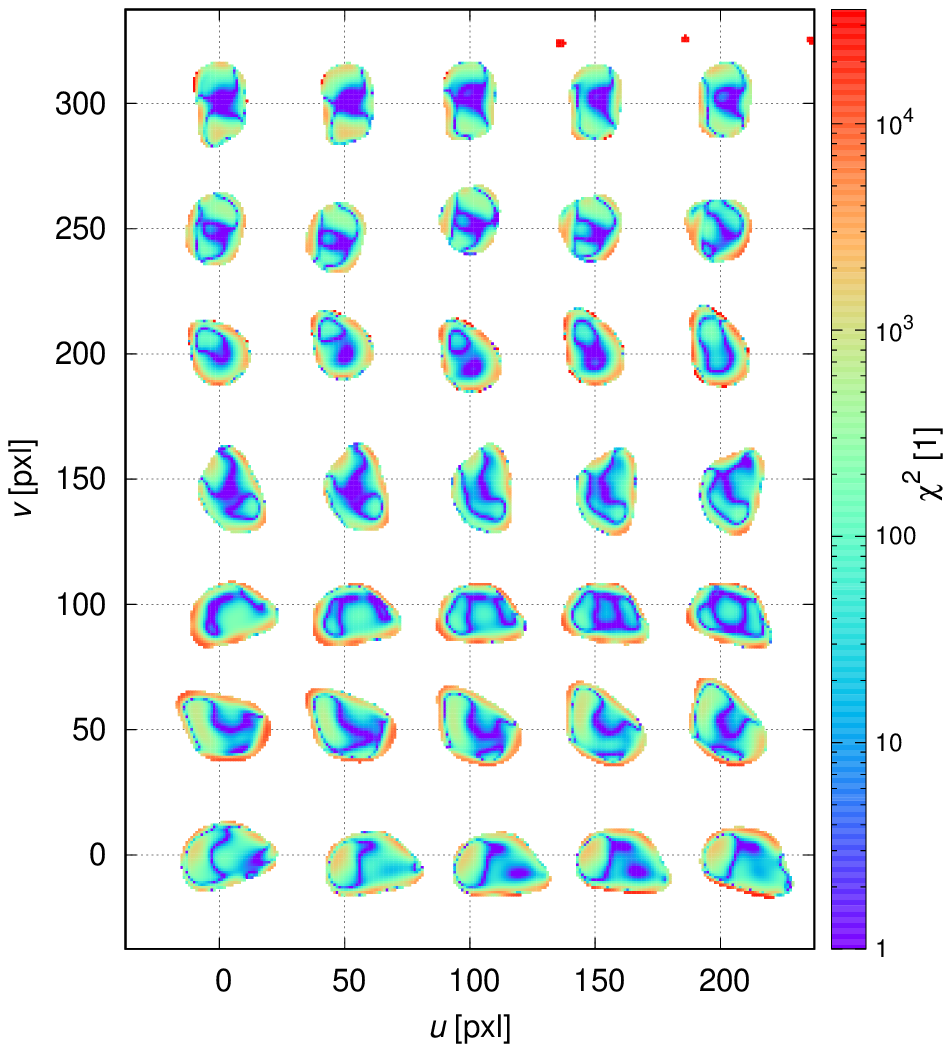}
\caption{
The residuals from fitting of 35 deconvolved AO images of (22) Kalliope
(taken from \citealt{Ferrais_2022A&A...662A..71F}),
with contributions to $\chi^2$ of individual pixels plotted in colour.
The projected shape changes due to rotation and viewing geometry.
The pixel scale is $3.6\,{\rm mas}\,{\rm pxl}^{-1}$.
After convergence of the shape parameters,
with help of the polygonal algorithm and `cliptracing',
the total $\chi^2_{\rm ao2} = 16684517$, $n_{\rm ao2} = 22843$.
Remaining systematics are partly due to rotation,
which changes the projected shape during an exposure in a non-trivial way.
Observed image of Linus was not fitted
(cf. red dots in the upper right corner).
}
\label{22_test54_aoellipsoid__23663_chi2_AO2}
\end{figure}


\section{Conclusions}

In this work, mutual occultation, transit and eclipse events
of Linus orbiting (22) Kalliope were used to constrain
combined astrometric + photometric models of this binary system.
Using innovative algorithms for photometric computations
(see Sects.~\ref{polygon}, \ref{cliptracing}),
we confirmed the size of Linus $(28\pm 1)\,{\rm km}$,
improved the shape of (22) Kalliope,
and put strong constraints on its dynamical oblateness. 

On one hand, we were not surprised by the low-oblateness ($C_{20} \simeq -0.12$) solution,
because (22) Kalliope is probably the best candidate
for a differentiated body \citep{Vernazza_2021A_A...654A..56V}.
At the same time, the iron core is usually considered to be more-or-less spherical,
because this is a standard outcome of differentiation.

\begin{figure}
\centering
\includegraphics[width=7cm]{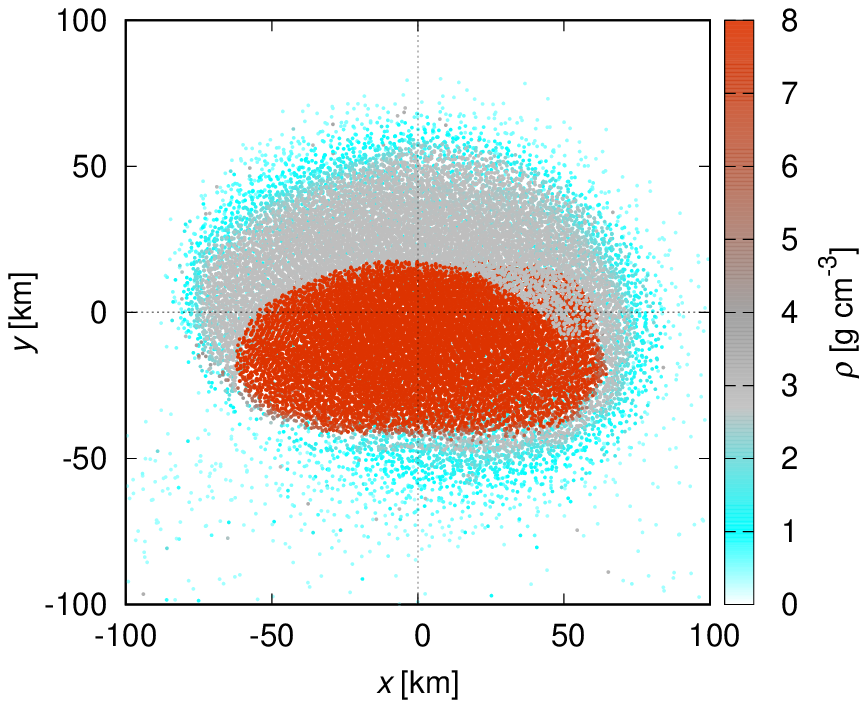}
\caption{
One of the SPH simulations from \cite{Broz_2022A&A...664A..69B},
showing a medium-energy impact to a differentiated body,
which created an elongated iron core.
The time 10000\,s corresponds to the end of fragmentation phase.
The density is indicated by the colour scale.
The oblateness $J_2 = -C_{20}$ is higher (not lower) than for a homogeneous body.
This could correspond to our long-arc, astrometric model
with 3 precession cycles.
}
\label{kalliope9_volume}
\end{figure}

On the other hand, we were surprised by the second, high-oblateness ($C_{20} \simeq -0.20$) solution.
Surprisingly, it should correspond to an irregular (or highly ellipsoidal) iron core.
In fact, (22) Kalliope suffered a major collision about 900\,My ago,
which gave birth to the Kalliope family \citep{Broz_2022A&A...664A..69B}.
Some of the SPH simulations of this event,
we performed in our previous work,
indicate that the iron core of the original body
is deformed and even elongated,
especially in medium- to high-energy collisions
(with the projectile size ${\gtrsim}\,45\,{\rm km}$).
One example is shown in Fig.~\ref{kalliope9_volume}.

In the future,
it should be possible to distinguish these two solutions
by new astrometric (imaging or speckle-interferometric) observations,
obtained at a suitable phase(s) of the precession cycle
(seen, e.g., in Fig.~\ref{22_test48_update__6695_orbit1}).


\begin{acknowledgements}
This work has been supported by the Czech Science Foundation through grants
21-11058S (M.~Bro\v z),
20-08218S (J.~\v Durech, J.~Hanu\v s).
In this work, measurements from the BlueEye600 telescope,
supported by the Charles University, were used.
The TRAPPIST is a project funded by the Belgian Fonds (National) de la Recherche
Scientifique (F.R.S.-FNRS) under grant PDR T.0120.21.
J.~de~Wit and MIT gratefully acknowledge financial support from the Heising-Simons Foundation,
Dr.~and Mrs.~Colin Masson and Dr.~Peter A. Gilman
for Artemis, the first telescope of the SPECULOOS network situated in Tenerife, Spain.
We also thank an anonymous referee for constructive comments.
\end{acknowledgements}

\bibliographystyle{aa}
\bibliography{references}

\begin{thebibliography}{44}
\expandafter\ifx\csname natexlab\endcsname\relax\def\natexlab#1{#1}\fi

\bibitem[{{Bro{\v{z}}}(2017)}]{Broz_2017ApJS..230...19B}
{Bro{\v{z}}}, M. 2017, \apjs, 230, 19

\bibitem[{{Bro{\v{z}}} {et~al.}(2022{\natexlab{a}}){Bro{\v{z}}}, {Ferrais},
  {Vernazza}, {{\v{S}}eve{\v{c}}ek}, \& {Jutzi}}]{Broz_2022A&A...664A..69B}
{Bro{\v{z}}}, M., {Ferrais}, M., {Vernazza}, P., {{\v{S}}eve{\v{c}}ek}, P., \&
  {Jutzi}, M. 2022{\natexlab{a}}, \aap, 664, A69

\bibitem[{{Bro{\v{z}}} {et~al.}(2021){Bro{\v{z}}}, {Marchis}, {Jorda},
  {Hanu{\v{s}}}, {Vernazza}, {Ferrais}, {Vachier}, {Rambaux}, {Marsset},
  {Viikinkoski}, {Jehin}, {Benseguane}, {Podlewska-Gaca}, {Carry}, {Drouard},
  {Fauvaud}, {Birlan}, {Berthier}, {Bartczak}, {Dumas}, {Dudzi{\'n}ski},
  {{\v{D}}urech}, {Castillo-Rogez}, {Cipriani}, {Colas}, {Fetick}, {Fusco},
  {Grice}, {Kryszczynska}, {Lamy}, {Marciniak}, {Michalowski}, {Michel},
  {Pajuelo}, {Santana-Ros}, {Tanga}, {Vigan}, {Vokrouhlick{\'y}}, {Witasse}, \&
  {Yang}}]{Broz_2021A&A...653A..56B}
{Bro{\v{z}}}, M., {Marchis}, F., {Jorda}, L., {et~al.} 2021, \aap, 653, A56

\bibitem[{{Bro{\v{z}}} {et~al.}(2022{\natexlab{b}}){Bro{\v{z}}},
  {{\v{D}}urech}, {Carry}, {Vachier}, {Marchis}, {Hanu{\v{s}}}, {Jorda},
  {Vernazza}, {Vokrouhlick{\'y}}, {Walterov{\'a}}, \&
  {Behrend}}]{Broz_2022A&A...657A..76B}
{Bro{\v{z}}}, M., {{\v{D}}urech}, J., {Carry}, B., {et~al.} 2022{\natexlab{b}},
  \aap, 657, A76

\bibitem[{{Burdanov} {et~al.}(2022){Burdanov}, {de Wit}, {Gillon}, {Rebolo},
  {Sebastian}, {Alonso}, {Sohy}, {Niraula}, {Garcia}, {Barkaoui}, {Chinchilla},
  {Ducrot}, {Murray}, {Pedersen}, {Jehin}, {McCormac}, \&
  {Z{\'u}{\~n}iga-Fern{\'a}ndez}}]{Burdanov_2022PASP..134j5001B}
{Burdanov}, A.~Y., {de Wit}, J., {Gillon}, M., {et~al.} 2022, \pasp, 134,
  105001

\bibitem[{{Cheng} {et~al.}(2018){Cheng}, {Rivkin}, {Michel}, {Atchison},
  {Barnouin}, {Benner}, {Chabot}, {Ernst}, {Fahnestock}, {Kueppers}, {Pravec},
  {Rainey}, {Richardson}, {Stickle}, \& {Thomas}}]{Cheng_2018P&SS..157..104C}
{Cheng}, A.~F., {Rivkin}, A.~S., {Michel}, P., {et~al.} 2018, \planss, 157, 104

\bibitem[{{Delrez} {et~al.}(2018){Delrez}, {Gillon}, {Queloz}, {Demory},
  {Almleaky}, {de Wit}, {Jehin}, {Triaud}, {Barkaoui}, {Burdanov}, {Burgasser},
  {Ducrot}, {McCormac}, {Murray}, {Silva Fernandes}, {Sohy}, {Thompson}, {Van
  Grootel}, {Alonso}, {Benkhaldoun}, \& {Rebolo}}]{Delrez_2018SPIE10700E..1ID}
{Delrez}, L., {Gillon}, M., {Queloz}, D., {et~al.} 2018, in Society of
  Photo-Optical Instrumentation Engineers (SPIE) Conference Series, Vol. 10700,
  Ground-based and Airborne Telescopes VII, ed. H.~K. {Marshall} \&
  J.~{Spyromilio}, 107001I

\bibitem[{{DeMeo} {et~al.}(2009){DeMeo}, {Binzel}, {Slivan}, \&
  {Bus}}]{Demeo_2009Icar..202..160D}
{DeMeo}, F.~E., {Binzel}, R.~P., {Slivan}, S.~M., \& {Bus}, S.~J. 2009,
  \icarus, 202, 160

\bibitem[{{Descamps} {et~al.}(2008){Descamps}, {Marchis}, {Pollock},
  {Berthier}, {Vachier}, {Birlan}, {Kaasalainen}, {Harris}, {Wong},
  {Romanishin}, {Cooper}, {Kettner}, {Wiggins}, {Kryszczynska}, {Polinska},
  {Coliac}, {Devyatkin}, {Verestchagina}, \&
  {Gorshanov}}]{Descamps_2008Icar..196..578D}
{Descamps}, P., {Marchis}, F., {Pollock}, J., {et~al.} 2008, \icarus, 196, 578

\bibitem[{{\v Durech} {et~al.}(2018){\v Durech}, {Hanu{\v{s}}}, {Bro{\v{z}}},
  {Lehk{\'y}}, {Behrend}, {Antonini}, {Charbonnel}, {Crippa}, {Dubreuil},
  {Farroni}, {Kober}, {Lopez}, {Manzini}, {Oey}, {Poncy}, {Rinner}, \&
  {Roy}}]{Durech_2018Icar..304..101D}
{\v Durech}, J., {Hanu{\v{s}}}, J., {Bro{\v{z}}}, M., {et~al.} 2018, \icarus,
  304, 101

\bibitem[{{\v Durech} {et~al.}(2010){\v Durech}, {Sidorin}, \&
  {Kaasalainen}}]{Durech_2010A&A...513A..46D}
{\v Durech}, J., {Sidorin}, V., \& {Kaasalainen}, M. 2010, \aap, 513, A46

\bibitem[{{Ferrais} {et~al.}(2022){Ferrais}, {Jorda}, {Vernazza}, {Carry},
  {Bro{\v{z}}}, {Rambaux}, {Hanu{\v{s}}}, {Dudzi{\'n}ski}, {Bartczak},
  {Vachier}, {Aristidi}, {Beck}, {Marchis}, {Marsset}, {Viikinkoski}, {Fetick},
  {Drouard}, {Fusco}, {Birlan}, {Podlewska-Gaca}, {Burbine}, {Dyar},
  {Bendjoya}, {Benkhaldoun}, {Berthier}, {Castillo-Rogez}, {Cipriani}, {Colas},
  {Dumas}, {{\v{D}}urech}, {Fauvaud}, {Grice}, {Jehin}, {Kaasalainen},
  {Kryszczynska}, {Lamy}, {Le Coroller}, {Marciniak}, {Michalowski}, {Michel},
  {Prieur}, {Reddy}, {Rivet}, {Santana-Ros}, {Scardia}, {Tanga}, {Vigan},
  {Witasse}, \& {Yang}}]{Ferrais_2022A&A...662A..71F}
{Ferrais}, M., {Jorda}, L., {Vernazza}, P., {et~al.} 2022, \aap, 662, A71

\bibitem[{{Gaia Collaboration} {et~al.}(2018){Gaia Collaboration}, {Brown},
  {Vallenari}, {Prusti}, {de Bruijne}, {Babusiaux}, {Bailer-Jones}, {Biermann},
  {Evans}, {Eyer}, {Jansen}, {Jordi}, {Klioner}, {Lammers}, {Lindegren},
  {Luri}, {Mignard}, {Panem}, {Pourbaix}, {Randich}, {Sartoretti}, {Siddiqui},
  {Soubiran}, {van Leeuwen}, {Walton}, {Arenou}, {Bastian}, {Cropper},
  {Drimmel}, {Katz}, {Lattanzi}, {Bakker}, {Cacciari}, {Casta{\~n}eda},
  {Chaoul}, {Cheek}, {De Angeli}, {Fabricius}, {Guerra}, {Holl}, {Masana},
  {Messineo}, {Mowlavi}, {Nienartowicz}, {Panuzzo}, {Portell}, {Riello},
  {Seabroke}, {Tanga}, {Th{\'e}venin}, {Gracia-Abril}, {Comoretto},
  {Garcia-Reinaldos}, {Teyssier}, {Altmann}, {Andrae}, {Audard},
  {Bellas-Velidis}, {Benson}, {Berthier}, {Blomme}, {Burgess}, {Busso},
  {Carry}, {Cellino}, {Clementini}, {Clotet}, {Creevey}, {Davidson}, {De
  Ridder}, {Delchambre}, {Dell'Oro}, {Ducourant},
  {Fern{\'a}ndez-Hern{\'a}ndez}, {Fouesneau}, {Fr{\'e}mat}, {Galluccio},
  {Garc{\'\i}a-Torres}, {Gonz{\'a}lez-N{\'u}{\~n}ez}, {Gonz{\'a}lez-Vidal},
  {Gosset}, {Guy}, {Halbwachs}, {Hambly}, {Harrison}, {Hern{\'a}ndez},
  {Hestroffer}, {Hodgkin}, {Hutton}, {Jasniewicz}, {Jean-Antoine-Piccolo},
  {Jordan}, {Korn}, {Krone-Martins}, {Lanzafame}, {Lebzelter}, {L{\"o}ffler},
  {Manteiga}, {Marrese}, {Mart{\'\i}n-Fleitas}, {Moitinho}, {Mora}, {Muinonen},
  {Osinde}, {Pancino}, {Pauwels}, {Petit}, {Recio-Blanco}, {Richards},
  {Rimoldini}, {Robin}, {Sarro}, {Siopis}, {Smith}, {Sozzetti}, {S{\"u}veges},
  {Torra}, {van Reeven}, {Abbas}, {Abreu Aramburu}, {Accart}, {Aerts},
  {Altavilla}, {{\'A}lvarez}, {Alvarez}, {Alves}, {Anderson}, {Andrei},
  {Anglada Varela}, {Antiche}, {Antoja}, {Arcay}, {Astraatmadja}, {Bach},
  {Baker}, {Balaguer-N{\'u}{\~n}ez}, {Balm}, {Barache}, {Barata}, {Barbato},
  {Barblan}, {Barklem}, {Barrado}, {Barros}, {Barstow}, {Bartholom{\'e}
  Mu{\~n}oz}, {Bassilana}, {Becciani}, {Bellazzini}, {Berihuete}, {Bertone},
  {Bianchi}, {Bienaym{\'e}}, {Blanco-Cuaresma}, {Boch}, {Boeche}, {Bombrun},
  {Borrachero}, {Bossini}, {Bouquillon}, {Bourda}, {Bragaglia}, {Bramante},
  {Breddels}, {Bressan}, {Brouillet}, {Br{\"u}semeister}, {Brugaletta},
  {Bucciarelli}, {Burlacu}, {Busonero}, {Butkevich}, {Buzzi}, {Caffau},
  {Cancelliere}, {Cannizzaro}, {Cantat-Gaudin}, {Carballo}, {Carlucci},
  {Carrasco}, {Casamiquela}, {Castellani}, {Castro-Ginard}, {Charlot},
  {Chemin}, {Chiavassa}, {Cocozza}, {Costigan}, {Cowell}, {Crifo}, {Crosta},
  {Crowley}, {Cuypers}, {Dafonte}, {Damerdji}, {Dapergolas}, {David}, {David},
  {de Laverny}, {De Luise}, {De March}, {de Martino}, {de Souza}, {de Torres},
  {Debosscher}, {del Pozo}, {Delbo}, {Delgado}, {Delgado}, {Di Matteo},
  {Diakite}, {Diener}, {Distefano}, {Dolding}, {Drazinos}, {Dur{\'a}n},
  {Edvardsson}, {Enke}, {Eriksson}, {Esquej}, {Eynard Bontemps}, {Fabre},
  {Fabrizio}, {Faigler}, {Falc{\~a}o}, {Farr{\`a}s Casas}, {Federici},
  {Fedorets}, {Fernique}, {Figueras}, {Filippi}, {Findeisen}, {Fonti},
  {Fraile}, {Fraser}, {Fr{\'e}zouls}, {Gai}, {Galleti}, {Garabato},
  {Garc{\'\i}a-Sedano}, {Garofalo}, {Garralda}, {Gavel}, {Gavras}, {Gerssen},
  {Geyer}, {Giacobbe}, {Gilmore}, {Girona}, {Giuffrida}, {Glass}, {Gomes},
  {Granvik}, {Gueguen}, {Guerrier}, {Guiraud}, {Guti{\'e}rrez-S{\'a}nchez},
  {Haigron}, {Hatzidimitriou}, {Hauser}, {Haywood}, {Heiter}, {Helmi}, {Heu},
  {Hilger}, {Hobbs}, {Hofmann}, {Holland}, {Huckle}, {Hypki}, {Icardi},
  {Jan{\ss}en}, {Jevardat de Fombelle}, {Jonker}, {Juh{\'a}sz}, {Julbe},
  {Karampelas}, {Kewley}, {Klar}, {Kochoska}, {Kohley}, {Kolenberg},
  {Kontizas}, {Kontizas}, {Koposov}, {Kordopatis}, {Kostrzewa-Rutkowska},
  {Koubsky}, {Lambert}, {Lanza}, {Lasne}, {Lavigne}, {Le Fustec}, {Le
  Poncin-Lafitte}, {Lebreton}, {Leccia}, {Leclerc}, {Lecoeur-Taibi},
  {Lenhardt}, {Leroux}, {Liao}, {Licata}, {Lindstr{\o}m}, {Lister}, {Livanou},
  {Lobel}, {L{\'o}pez}, {Managau}, {Mann}, {Mantelet}, {Marchal}, {Marchant},
  {Marconi}, {Marinoni}, {Marschalk{\'o}}, {Marshall}, {Martino}, {Marton},
  {Mary}, {Massari}, {Matijevi{\v{c}}}, {Mazeh}, {McMillan}, {Messina},
  {Michalik}, {Millar}, {Molina}, {Molinaro}, {Moln{\'a}r}, {Montegriffo},
  {Mor}, {Morbidelli}, {Morel}, {Morris}, {Mulone}, {Muraveva}, {Musella},
  {Nelemans}, {Nicastro}, {Noval}, {O'Mullane}, {Ord{\'e}novic},
  {Ord{\'o}{\~n}ez-Blanco}, {Osborne}, {Pagani}, {Pagano}, {Pailler},
  {Palacin}, {Palaversa}, {Panahi}, {Pawlak}, {Piersimoni}, {Pineau}, {Plachy},
  {Plum}, {Poggio}, {Poujoulet}, {Pr{\v{s}}a}, {Pulone}, {Racero}, {Ragaini},
  {Rambaux}, {Ramos-Lerate}, {Regibo}, {Reyl{\'e}}, {Riclet}, {Ripepi}, {Riva},
  {Rivard}, {Rixon}, {Roegiers}, {Roelens}, {Romero-G{\'o}mez}, {Rowell},
  {Royer}, {Ruiz-Dern}, {Sadowski}, {Sagrist{\`a} Sell{\'e}s}, {Sahlmann},
  {Salgado}, {Salguero}, {Sanna}, {Santana-Ros}, {Sarasso}, {Savietto},
  {Schultheis}, {Sciacca}, {Segol}, {Segovia}, {S{\'e}gransan}, {Shih},
  {Siltala}, {Silva}, {Smart}, {Smith}, {Solano}, {Solitro}, {Sordo}, {Soria
  Nieto}, {Souchay}, {Spagna}, {Spoto}, {Stampa}, {Steele},
  {Steidelm{\"u}ller}, {Stephenson}, {Stoev}, {Suess}, {Surdej}, {Szabados},
  {Szegedi-Elek}, {Tapiador}, {Taris}, {Tauran}, {Taylor}, {Teixeira},
  {Terrett}, {Teyssandier}, {Thuillot}, {Titarenko}, {Torra Clotet}, {Turon},
  {Ulla}, {Utrilla}, {Uzzi}, {Vaillant}, {Valentini}, {Valette}, {van Elteren},
  {Van Hemelryck}, {van Leeuwen}, {Vaschetto}, {Vecchiato}, {Veljanoski},
  {Viala}, {Vicente}, {Vogt}, {von Essen}, {Voss}, {Votruba}, {Voutsinas},
  {Walmsley}, {Weiler}, {Wertz}, {Wevers}, {Wyrzykowski}, {Yoldas},
  {{\v{Z}}erjal}, {Ziaeepour}, {Zorec}, {Zschocke}, {Zucker}, {Zurbach}, \&
  {Zwitter}}]{Brown_2018A&A...616A...1G}
{Gaia Collaboration}, {Brown}, A.~G.~A., {Vallenari}, A., {et~al.} 2018, \aap,
  616, A1

\bibitem[{{Gehrels} \& {Owings}(1962)}]{Gehrels_1962ApJ...135..906G}
{Gehrels}, T. \& {Owings}, D. 1962, \apj, 135, 906

\bibitem[{{Hanu{\v{s}}} {et~al.}(2016){Hanu{\v{s}}}, {{\v{D}}urech},
  {Oszkiewicz}, {Behrend}, {Carry}, {Delbo}, {Adam}, {Afonina}, {Anquetin},
  {Antonini}, {Arnold}, {Audejean}, {Aurard}, {Bachschmidt}, {Baduel},
  {Barbotin}, {Barroy}, {Baudouin}, {Berard}, {Berger}, {Bernasconi}, {Bosch},
  {Bouley}, {Bozhinova}, {Brinsfield}, {Brunetto}, {Canaud}, {Caron},
  {Carrier}, {Casalnuovo}, {Casulli}, {Cerda}, {Chalamet}, {Charbonnel},
  {Chinaglia}, {Cikota}, {Colas}, {Coliac}, {Collet}, {Coloma}, {Conjat},
  {Conseil}, {Costa}, {Crippa}, {Cristofanelli}, {Damerdji}, {Deback{\`e}re},
  {Decock}, {D{\'e}hais}, {D{\'e}l{\'e}age}, {Delmelle}, {Demeautis},
  {Dr{\'o}{\.z}d{\.z}}, {Dubos}, {Dulcamara}, {Dumont}, {Durkee}, {Dymock},
  {Escalante del Valle}, {Esseiva}, {Esseiva}, {Esteban}, {Fauchez},
  {Fauerbach}, {Fauvaud}, {Fauvaud}, {Forn{\'e}}, {Fournel}, {Fradet},
  {Garlitz}, {Gerteis}, {Gillier}, {Gillon}, {Giraud}, {Godard}, {Goncalves},
  {Hamanowa}, {Hamanowa}, {Hay}, {Hellmich}, {Heterier}, {Higgins}, {Hirsch},
  {Hodosan}, {Hren}, {Hygate}, {Innocent}, {Jacquinot}, {Jawahar}, {Jehin},
  {Jerosimic}, {Klotz}, {Koff}, {Korlevic}, {Kosturkiewicz}, {Krafft},
  {Krugly}, {Kugel}, {Labrevoir}, {Lecacheux}, {Lehk{\'y}}, {Leroy},
  {Lesquerbault}, {Lopez-Gonzales}, {Lutz}, {Mallecot}, {Manfroid}, {Manzini},
  {Marciniak}, {Martin}, {Modave}, {Montaigut}, {Montier}, {Morelle}, {Morton},
  {Mottola}, {Naves}, {Nomen}, {Oey}, {Og{\l}oza}, {Paiella}, {Pallares},
  {Peyrot}, {Pilcher}, {Pirenne}, {Piron}, {Poli{\'n}ska}, {Polotto}, {Poncy},
  {Previt}, {Reignier}, {Renauld}, {Ricci}, {Richard}, {Rinner}, {Risoldi},
  {Robilliard}, {Romeuf}, {Rousseau}, {Roy}, {Ruthroff}, {Salom}, {Salvador},
  {Sanchez}, {Santana-Ros}, {Scholz}, {S{\'e}n{\'e}}, {Skiff}, {Sobkowiak},
  {Sogorb}, {Sold{\'a}n}, {Spiridakis}, {Splanska}, {Sposetti}, {Starkey},
  {Stephens}, {Stiepen}, {Stoss}, {Strajnic}, {Teng}, {Tumolo}, {Vagnozzi},
  {Vanoutryve}, {Vugnon}, {Warner}, {Waucomont}, {Wertz}, {Winiarski}, \&
  {Wolf}}]{Hanus_2016A_A...586A.108H}
{Hanu{\v{s}}}, J., {{\v{D}}urech}, J., {Oszkiewicz}, D.~A., {et~al.} 2016,
  \aap, 586, A108

\bibitem[{{Hapke}(1981)}]{Hapke_1981JGR....86.3039H}
{Hapke}, B. 1981, \jgr, 86, 3039

\bibitem[{{Herald} {et~al.}(2019){Herald}, {Frappa}, {Gault}, {Hayamizu},
  {Kerr}, {Moore}, \& {Giacchini}}]{Herald_2019pdss.data....3H}
{Herald}, D., {Frappa}, E., {Gault}, D., {et~al.} 2019, NASA Planetary Data
  System, 3

\bibitem[{{Herald} {et~al.}(2020){Herald}, {Gault}, {Anderson}, {Dunham},
  {Frappa}, {Hayamizu}, {Kerr}, {Miyashita}, {Moore}, {Pavlov}, {Preston},
  {Talbot}, \& {Timerson}}]{Herald_2020MNRAS.499.4570H}
{Herald}, D., {Gault}, D., {Anderson}, R., {et~al.} 2020, \mnras, 499, 4570

\bibitem[{{IAU SOFA Center}(2014)}]{SOFA_2014ascl.soft03026I}
{IAU SOFA Center}. 2014, {SOFA: Standards of Fundamental Astronomy},
  Astrophysics Source Code Library, record ascl:1403.026

\bibitem[{{Jehin} {et~al.}(2011){Jehin}, {Gillon}, {Queloz}, {Magain},
  {Manfroid}, {Chantry}, {Lendl}, {Hutsem{\'e}kers}, \&
  {Udry}}]{Jehin_2011Msngr.145....2J}
{Jehin}, E., {Gillon}, M., {Queloz}, D., {et~al.} 2011, The Messenger, 145, 2

\bibitem[{{Kobbelt}(2000)}]{Kobbelt_2000}
{Kobbelt}, L. 2000, {Proc. Computer graphics and interactive techniques}, 103

\bibitem[{{Li} {et~al.}(2015){Li}, {Helfenstein}, {Buratti}, {Takir}, \&
  {Clark}}]{Li_2015aste.book..129L}
{Li}, J.~Y., {Helfenstein}, P., {Buratti}, B., {Takir}, D., \& {Clark}, B.~E.
  2015, in Asteroids IV, 129--150

\bibitem[{{Lieske} {et~al.}(1977){Lieske}, {Lederle}, {Fricke}, \&
  {Morando}}]{Lieske_1977A&A....58....1L}
{Lieske}, J.~H., {Lederle}, T., {Fricke}, W., \& {Morando}, B. 1977, \aap, 58,
  1

\bibitem[{{Lupishko} {et~al.}(1982){Lupishko}, {Belskaia}, {Tupieva}, \&
  {Chernova}}]{Lupishko_1982AVest..16..101L}
{Lupishko}, D.~F., {Belskaia}, I.~N., {Tupieva}, F.~A., \& {Chernova}, G.~P.
  1982, Astronomicheskii Vestnik, 16, 101

\bibitem[{{Park} {et~al.}(2018){Park}, {Yim}, {Choi}, {Jo}, {Moon}, {Park},
  {Bae}, {Park}, {Roh}, {Cho}, {Choi}, {Kim}, \&
  {Choi}}]{Park_2018AdSpR..62..152P}
{Park}, J.-H., {Yim}, H.-S., {Choi}, Y.-J., {et~al.} 2018, Advances in Space
  Research, 62, 152

\bibitem[{{Pravec} \& {Hahn}(1997)}]{Pravec_1997Icar..127..431P}
{Pravec}, P. \& {Hahn}, G. 1997, \icarus, 127, 431

\bibitem[{{Pr{\v{s}}a} {et~al.}(2016){Pr{\v{s}}a}, {Conroy}, {Horvat}, {Pablo},
  {Kochoska}, {Bloemen}, {Giammarco}, {Hambleton}, \&
  {Degroote}}]{Prsa_2016ApJS..227...29P}
{Pr{\v{s}}a}, A., {Conroy}, K.~E., {Horvat}, M., {et~al.} 2016, \apjs, 227, 29

\bibitem[{{Ragozzine} \& {Brown}(2009)}]{Ragozzine_2009AJ....137.4766R}
{Ragozzine}, D. \& {Brown}, M.~E. 2009, \aj, 137, 4766

\bibitem[{{Rowan}(1990)}]{Rowan_1990}
{Rowan}, N. 1990, Ph.D. thesis, Univ. Texas Austin

\bibitem[{{Scaltriti} {et~al.}(1978){Scaltriti}, {Zappala}, \&
  {Stanzel}}]{Scaltriti_1978Icar...34...93S}
{Scaltriti}, F., {Zappala}, V., \& {Stanzel}, R. 1978, \icarus, 34, 93

\bibitem[{{Scardia} {et~al.}(2019){Scardia}, {Rivet}, {Prieur}, {Pansecchi},
  {Argyle}, {Ling}, {Aristidi}, {Zanutta}, {Vernet}, {Abe}, {Bendjoya},
  {Dimur}, \& {Su{\'a}rez}}]{Scardia_2019AN....340..771S}
{Scardia}, M., {Rivet}, J.-P., {Prieur}, J.-L., {et~al.} 2019, Astronomische
  Nachrichten, 340, 771

\bibitem[{{Scheirich} \& {Pravec}(2022)}]{Scheirich_2022PSJ.....3..163S}
{Scheirich}, P. \& {Pravec}, P. 2022, \psj, 3, 163

\bibitem[{{Spjuth}(2009)}]{Spjuth_2009PhDT.......588S}
{Spjuth}, S. 2009, PhD thesis, Technical University of Braunschweig, Germany

\bibitem[{{Statler} {et~al.}(2022){Statler}, {Raducan}, {Barnouin}, {DeCoster},
  {Chesley}, {Barbee}, {Agrusa}, {Cambioni}, {Cheng}, {Dotto}, {Eggl},
  {Fahnestock}, {Ferrari}, {Graninger}, {Herique}, {Herreros}, {Hirabayashi},
  {Ivanovski}, {Jutzi}, {Karatekin}, {Lucchetti}, {Luther}, {Makadia},
  {Marzari}, {Michel}, {Murdoch}, {Nakano}, {Orm{\"o}}, {Pajola}, {Rivkin},
  {Rossi}, {S{\'a}nchez}, {Schwartz}, {Soldini}, {Souami}, {Stickle},
  {Tortora}, {Trigo-Rodr{\'\i}guez}, {Venditti}, {Vincent}, \&
  {W{\"u}nnemann}}]{Statler_2022PSJ.....3..244S}
{Statler}, T.~S., {Raducan}, S.~D., {Barnouin}, O.~S., {et~al.} 2022, \psj, 3,
  244

\bibitem[{{Sung} {et~al.}(2012){Sung}, {Park}, {Lee}, {Lee}, {Seong}, \&
  {Oh}}]{Sung_2012PKAS...27...95S}
{Sung}, H.-I., {Park}, Y.-H., {Lee}, S.-M., {et~al.} 2012, Publication of
  Korean Astronomical Society, 27, 95

\bibitem[{{Surdej} {et~al.}(1986){Surdej}, {Pospieszalska-Surdej},
  {Michalowski}, \& {Schober}}]{Surdej_1986A&A...170..167S}
{Surdej}, J., {Pospieszalska-Surdej}, A., {Michalowski}, T., \& {Schober},
  H.~J. 1986, \aap, 170, 167

\bibitem[{{Thomas} \& {et al.}(2023)}]{Thomas_2023}
{Thomas}, C. \& {et al.} 2023, Nature, in press

\bibitem[{{van Leeuwen} {et~al.}(2018){van Leeuwen}, {de Bruijne}, {Arenou},
  {Bakker}, {Blomme}, {Busso}, {Cacciari}, {Casta{\~n}eda}, {Cellino},
  {Clotet}, {Comoretto}, {Eyer}, {Gonz{\'a}lez-N{\'u}{\~n}ez}, {Guy}, {Hambly},
  {Hobbs}, {van Leeuwen}, {Luri}, {Manteiga}, {Pourbaix}, {Roegiers},
  {Salgado}, {Sartoretti}, {Tanga}, {Ulla}, {Utrilla Molina}, {Abreu},
  {Altmann}, {Andrae}, {Antoja}, {Audard}, {Babusiaux}, {Bailer-Jones},
  {Barache}, {Bastian}, {Beck}, {Berthier}, {Bianchi}, {Biermann}, {Bombrun},
  {Bossini}, {Breddels}, {Brown}, {Busonero}, {Butkevich}, {Cantat-Gaudin},
  {Carrasco}, {Cheek}, {Clementini}, {Creevey}, {Crowley}, {David}, {Davidson},
  {De Angeli}, {De Ridder}, {Delb{\`o}}, {Dell'Oro}, {Diakit{\'e}},
  {Distefano}, {Drimmel}, {Dur{\'a}n}, {Evans}, {Fabricius}, {Fabrizio},
  {Fern{\'a}ndez-Hern{\'a}ndez}, {Findeisen}, {Fleitas}, {Fouesneau},
  {Galluccio}, {Gracia-Abril}, {Guerra}, {Guti{\'e}rrez-S{\'a}nchez}, {Helmi},
  {Hernandez}, {Holl}, {Hutton}, {Jean-Antoine-Piccolo}, {Jevardat de
  Fombelle}, {Joliet}, {Jordi}, {Juh{\'a}sz}, {Klioner}, {L{\"o}ffler},
  {Lammers}, {Lanzafame}, {Lebzelter}, {Leclerc}, {Lecoeur-Ta{\"\i}bi},
  {Lindegren}, {Marinoni}, {Marrese}, {Mary}, {Massari}, {Messineo},
  {Michalik}, {Mignard}, {Molinaro}, {Moln{\'a}r}, {Montegriffo}, {Mora},
  {Mowlavi}, {Muinonen}, {Muraveva}, {Nienartowicz}, {Ordenovic}, {Pancino},
  {Panem}, {Pauwels}, {Petit}, {Plachy}, {Portell}, {Racero}, {Regibo},
  {Reyl{\'e}}, {Rimoldini}, {Ripepi}, {Riva}, {Robichon}, {Robin}, {Roelens},
  {Romero-G{\'o}mez}, {Sarro}, {Seabroke}, {Segovia}, {Siddiqui}, {Smart},
  {Smith}, {Sordo}, {Soria}, {Spoto}, {Stephenson}, {Turon}, {Vallenari},
  {Veljanoski}, \& {Voutsinas}}]{Leeuwen_2018gdr2.reptE....V}
{van Leeuwen}, F., {de Bruijne}, J.~H.~J., {Arenou}, F., {et~al.} 2018, {Gaia
  DR2 documentation}, Gaia DR2 documentation, European Space Agency; Gaia Data
  Processing and Analysis Consortium.

\bibitem[{{Vatti}(1992)}]{Vatti_1992}
{Vatti}, B.~R. 1992, Comm.\ ACM, 35, 56

\bibitem[{{Vernazza} {et~al.}(2021){Vernazza}, {Ferrais}, {Jorda},
  {Hanu{\v{s}}}, {Carry}, {Marsset}, {Bro{\v{z}}}, {Fetick}, {Viikinkoski},
  {Marchis}, {Vachier}, {Drouard}, {Fusco}, {Birlan}, {Podlewska-Gaca},
  {Rambaux}, {Neveu}, {Bartczak}, {Dudzi{\'n}ski}, {Jehin}, {Beck}, {Berthier},
  {Castillo-Rogez}, {Cipriani}, {Colas}, {Dumas}, {{\v{D}}urech}, {Grice},
  {Kaasalainen}, {Kryszczynska}, {Lamy}, {Le Coroller}, {Marciniak},
  {Michalowski}, {Michel}, {Santana-Ros}, {Tanga}, {Vigan}, {Witasse}, {Yang},
  {Antonini}, {Audejean}, {Aurard}, {Behrend}, {Benkhaldoun}, {Bosch},
  {Chapman}, {Dalmon}, {Fauvaud}, {Hamanowa}, {Hamanowa}, {His}, {Jones},
  {Kim}, {Kim}, {Krajewski}, {Labrevoir}, {Leroy}, {Livet}, {Molina},
  {Montaigut}, {Oey}, {Payre}, {Reddy}, {Sabin}, {Sanchez}, \&
  {Socha}}]{Vernazza_2021A_A...654A..56V}
{Vernazza}, P., {Ferrais}, M., {Jorda}, L., {et~al.} 2021, \aap, 654, A56

\bibitem[{{Viikinkoski} {et~al.}(2015){Viikinkoski}, {Kaasalainen}, \&
  {Durech}}]{Viikinkoski_2015A&A...576A...8V}
{Viikinkoski}, M., {Kaasalainen}, M., \& {Durech}, J. 2015, \aap, 576, A8

\bibitem[{{Wahr}(1981)}]{Wahr_1981GeoJ...64..705W}
{Wahr}, J.~M. 1981, Geophysical Journal, 64, 705

\bibitem[{{Wolf}(1992)}]{Wolf_1992aspr.book.....W}
{Wolf}, M. 1992, {Astronomická příručka} ({Academia})

\bibitem[{{Wong} \& {Brown}(2019)}]{Wong_2019AJ....157..203W}
{Wong}, I. \& {Brown}, M.~E. 2019, \aj, 157, 203

\end{thebibliography}


\appendix
\section{Supplementary figures}

\begin{figure}
\centering
\includegraphics[width=9cm]{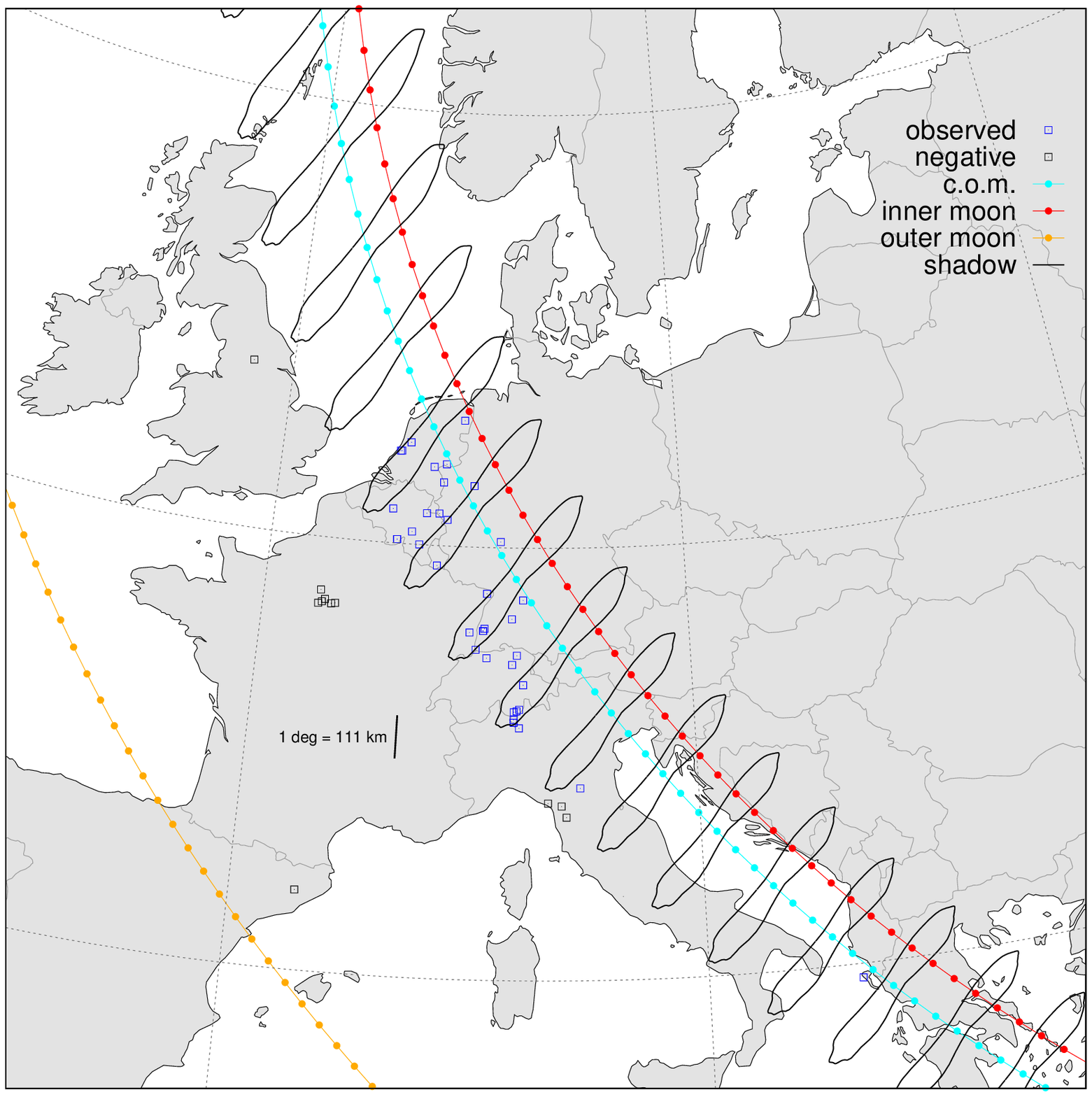}
\caption{
A verification of the stellar occultation algorithm
for the occultation of the star HIP\,54599 by asteroid (216) Kleopatra,
on Mar 12th 2015.
The shadow on the WGS-84 ellipsoid is plotted in black,
the centre-of-mass location as \color{cyan}cyan\color{black}\ line,
the 1st moon \color{red}red\color{black},
the 2nd moon \color{orange}orange\color{black},
positive measurements as \color{blue}blue\color{black}\ squares,
negative measurements \color{gray}gray\color{black}.
The gnomonic projection was used in this test.
}
\label{216_test79_20150312_occultation3}
\end{figure}

\begin{figure}
\centering
\includegraphics[width=8cm]{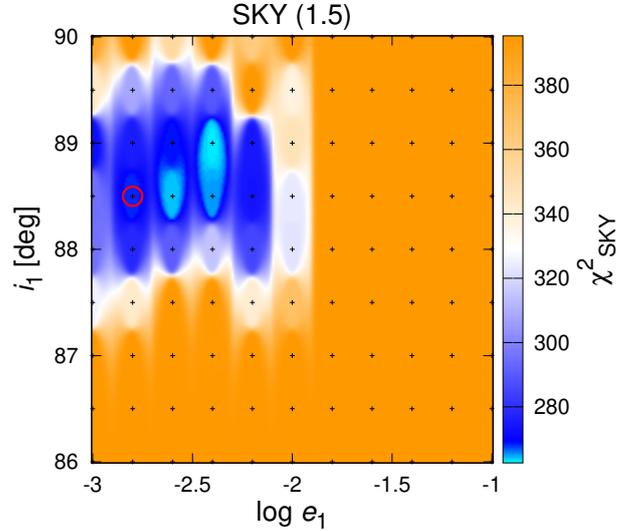}
\caption{
Similar as Fig.~\ref{22_fitting10_GRIDC20__249_C201_Deltat1_SKY}.
The logarithm of eccentricity $\log e_1$ vs. the inclination $i_1$ is plotted.
Models were converged for 99 combinations of the fixed parameters;
all other parameters were free.
For each combination, 1000 iteration were computed,
i.e., 99000 models in total.
The overall best-fit $\chi^2_{\rm sky} = 264$, $n_{\rm sky} = 344$,
indicated by \color{red}red\color{black}\ circle.
}
\label{22_fitting2_GRIDECC__310_loge1_i1_SKY}
\end{figure}

Stellar occultation algorithm verification for (216) Kleopatra
is show in Fig.~\ref{216_test79_20150312_occultation3}.
Additional models from Sec.~\ref{longarc_astrometric}
are shown in Fig.~\ref{22_fitting2_GRIDECC__310_loge1_i1_SKY}.
The corner plot discussed in Sec.~\ref{longarc_astrometric_photometric} si shown in
Fig.~\ref{22_test48_update__6695_corner_}. 
The adjusted shape model and the respective light curve fit
discussed in Sec.~\ref{shape} is shown in
Figs.~\ref{22_test51_ellipsoid__4195_shape},
\ref{22_test51_ellipsoid__4195_chi2_LC2_PHASE}.
Additional best-fit model is presented in Tab.~\ref{taba1}.

\begin{figure*}
\centering
\includegraphics[width=16cm]{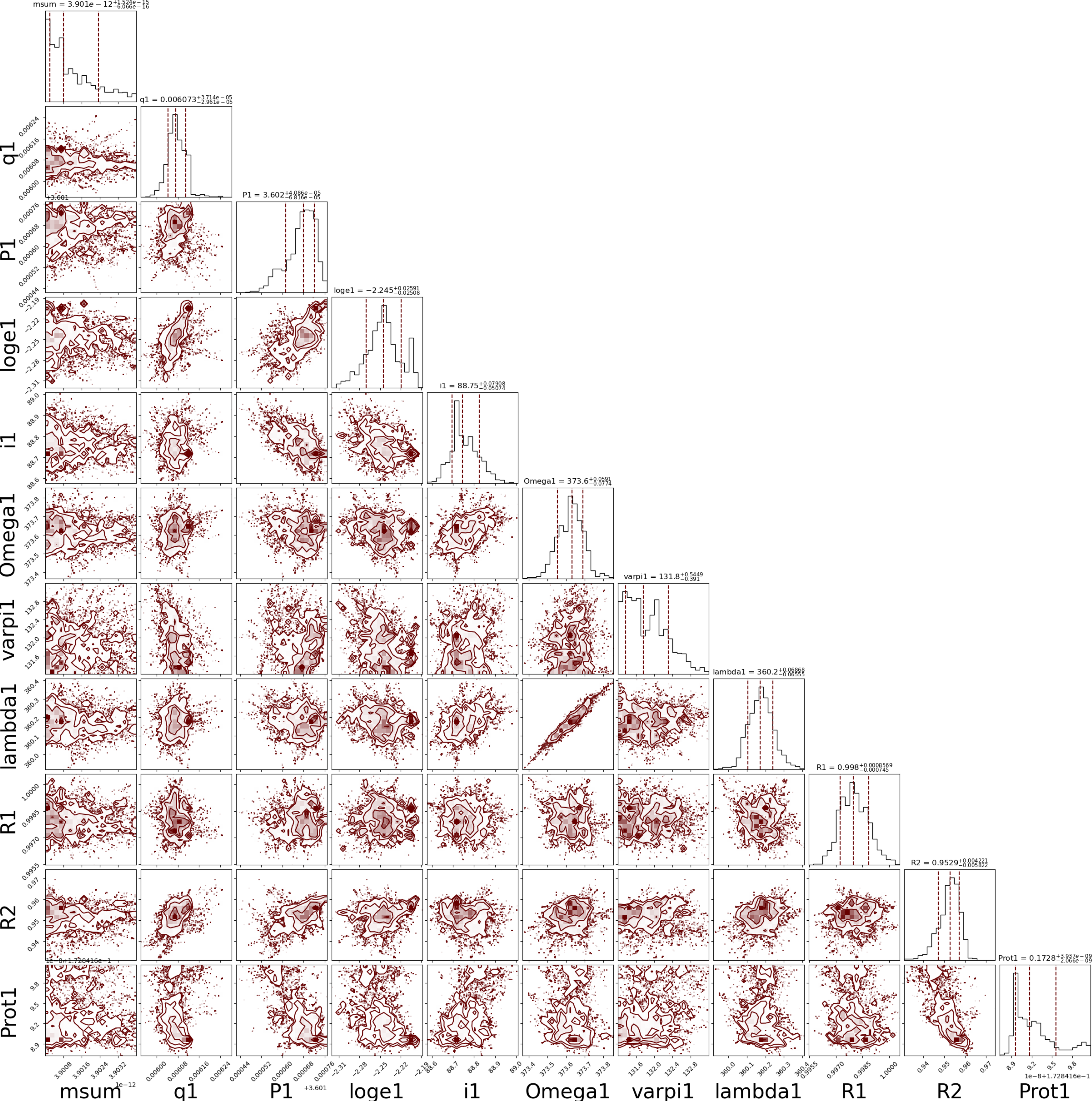}
\caption{
MCMC simulation for the model with $\chi^2 = 6695$ (from Tab.~\ref{tab2}).
Distributions of all parameters,
corresponding to local uncertainties,
and correlations of all pairs of parameters
is plotted as a standard `corner'.
The order of 22~parameters is as follows ($\downarrow$, $\rightarrow$):
$m_{\rm sum}$,
$q_1$,
$P_1$,
$\log e_1$,
$i_1$,
$\Omega_1$,
$\varpi_1$,
$\lambda_1$,
$R_1$,
$R_2$,
$P_{{\rm rot}1}$,
$\Delta t_1$,
$C_{20,1}$,
$l_{{\rm pole}1}$,
$b_{{\rm pole}1}$,
$\phi_{01}$,
$A_{{\rm w}1}$,
$A_{{\rm w}2}$,
$B_0$,
$h$,
$g$,
$\bar\theta$.
The mean value is seen in the respective histogram.
The number of walkers was set to 64.
The whole chain contained 2400 samples, the burn-in phase took up to 1000 of them.
In the course of iterations, walkers may drift away from the initial local minimum,
because walkers perform also low-probability steps to higher $\chi^2$ values,
which may sometimes result in systematic shifts of (some of) the parameters.
In this case, the MCMC was also affected by a change of the weights
($w_{\rm sky} = 10$, $w_{\rm ao} = 0.1$),
which prevents the MCMC from fitting the systematics on the light curve
at expense of the PISCO astrometric dataset.
}
\label{22_test48_update__6695_corner_}
\end{figure*}

\setcounter{figure}{2}
\begin{figure*}
\centering
\includegraphics[width=16cm]{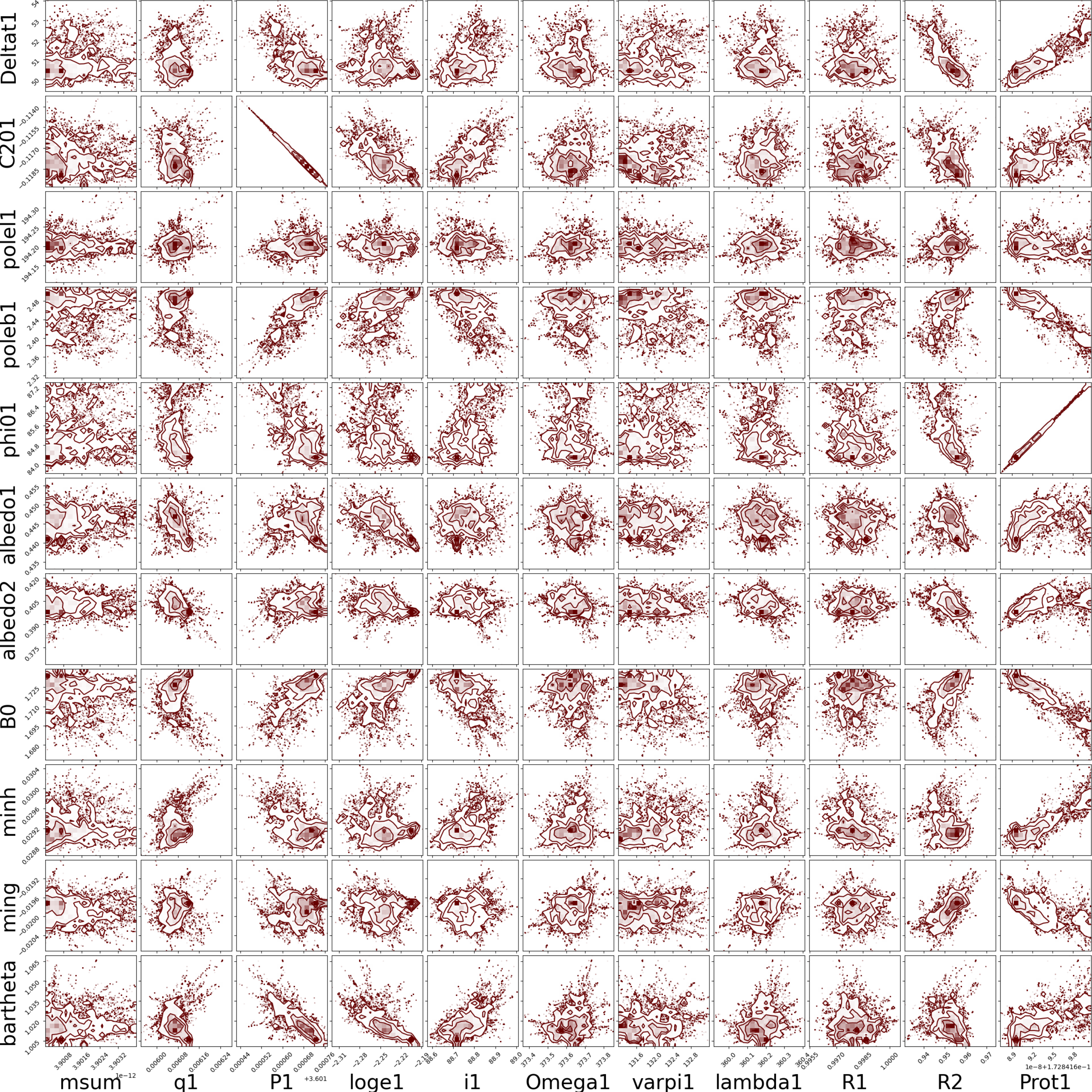}
\caption{(cont.)}
\end{figure*}

\setcounter{figure}{2}
\begin{figure*}
\centering
\includegraphics[width=16cm]{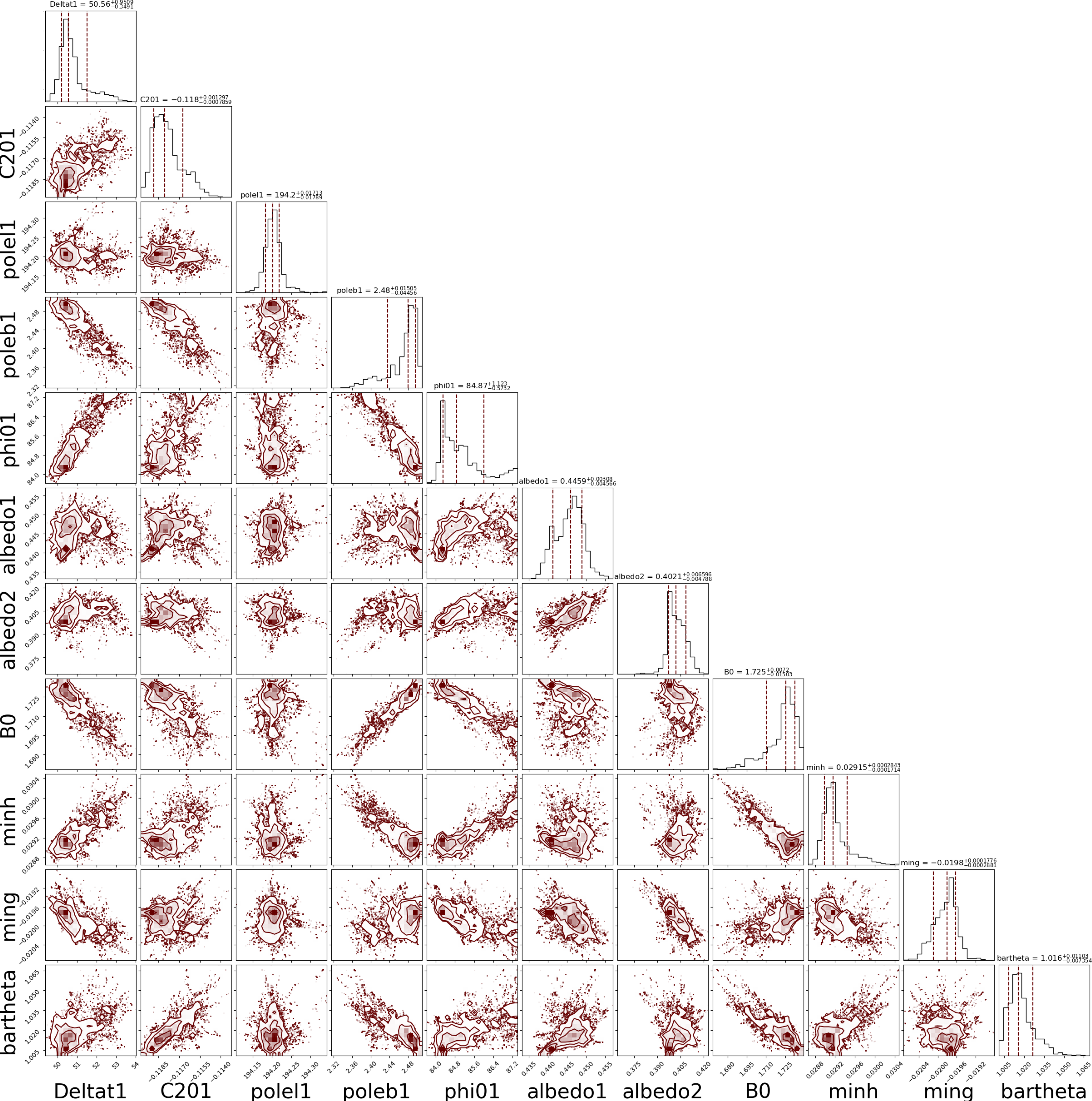}
\caption{(cont.)}
\end{figure*}

\begin{figure}
\centering
\includegraphics[width=9cm]{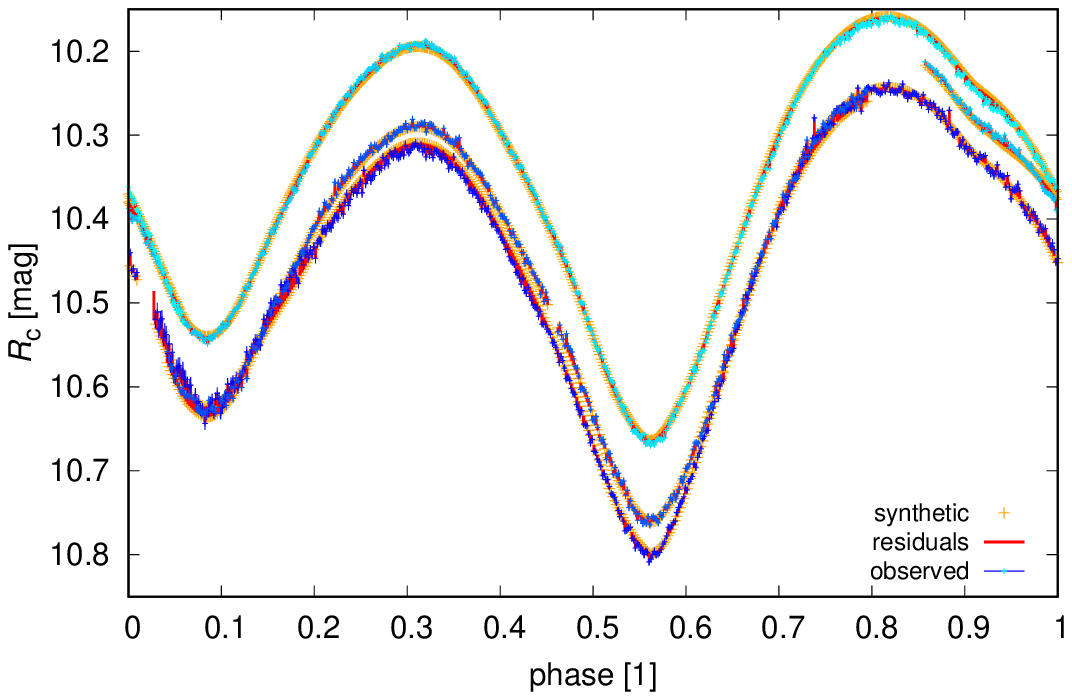}
\includegraphics[width=9cm]{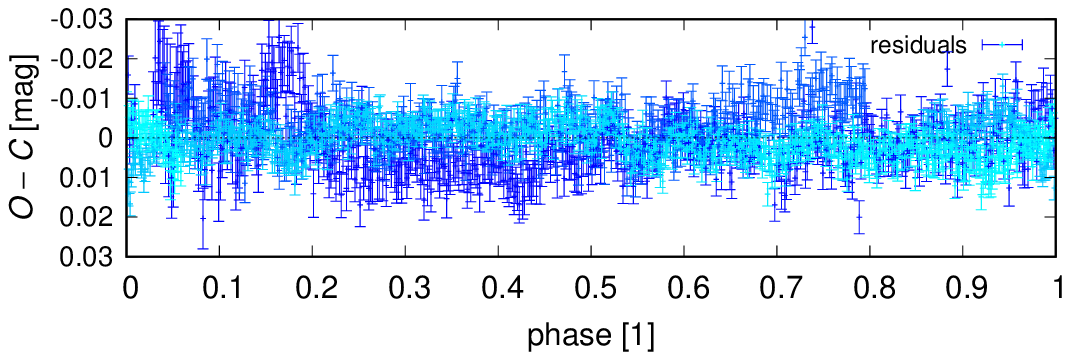}
\caption{
Same as Fig.~\ref{22_test25_shortarc__6462_chi2_LC2_PHASE},
but for the adjusted shape model of (22) Kalliope.
Systematics on the light curves related to the shape were at least partly eliminated.
The respective contribution decreased to
$\chi^2_{\rm lc} = 3980$, $n_{\rm lc} = 1829$.
}
\label{22_test51_ellipsoid__4195_chi2_LC2_PHASE}
\end{figure}

\begin{figure}
\centering
\includegraphics[width=6cm]{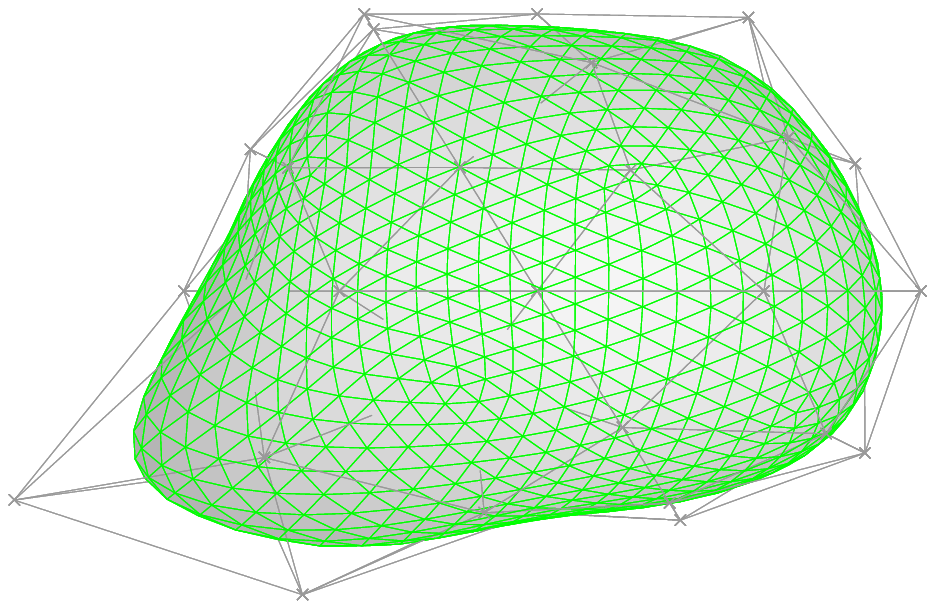}
\includegraphics[width=6cm]{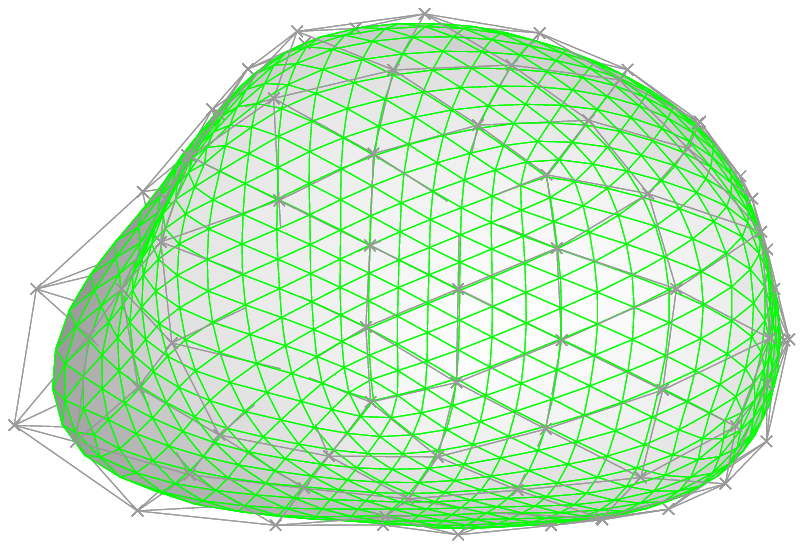}
\caption{
A comparison of the adjusted shape of (22) Kalliope (top)
vs. the original ADAM shape (bottom).
The control points (\color{gray}gray\color{black}) and 
the high-resolution mesh (\color{green}green\color{black}) are indicated.
Shades of gray correspond to the $z$ coordinate.
}
\label{22_test51_ellipsoid__4195_shape}
\end{figure}

\begin{table}
\caption{Same as Tab.~\ref{tab2} for the all-data model.}
\label{taba1}
\centering
\begin{tabular}{lll}
& all-data \\
var. & val. & unit \\
\hline
\vrule width 0pt height 10pt
$m_{\rm sum}$		& $3.902212\cdot 10^{-12}	$ & $M_{\rm S}$		\\
$q_1$			& $6.129\cdot 10^{-3}{}^{\rm f}	$ & 1			\\
$P_1$			& $3.601780    	 		$ & day			\\
$\log e_1$		& $-2.270     			$ & 1			\\
$i_1$			& $ 88.734     	 		$ & deg			\\
$\Omega_1$		& $374.913     	 		$ & deg			\\
$\varpi_1$		& $131.561     	 		$ & deg			\\
$\lambda_1$		& $360.703      		$ & deg			\\
$R_1$			& $0.987     			$ & 76.5\,km		\\
$R_2$			& $0.902     			$ & 15\,km		\\
$P_{\rm rot1}$		& $0.172841     		$ & day			\\
$P_{\rm rot2}$		& $3.595713^{\rm f}	   	$ & day			\\
$\Delta t_1$		& $49.9     		 	$ & s			\\
$C_{20,1}$		& $-0.1197    			$ & 1			\\
$l_{\rm pole1}$		& $195.010     	 		$ & deg			\\
$b_{\rm pole1}$		& $2.677	      		$ & deg			\\
$\phi_{01}$		& $84.577      			$ & deg			\\
$A_{\rm w1}$		& $0.442     			$ & 1			\\
$A_{\rm w2}$		& $0.303     			$ & 1			\\
$B_0$			& $1.733      			$ & 1			\\
$h$			& $0.0295 			$ & 1			\\
$g$			& $-0.0197			$ & 1			\\
$\bar\theta$		& $0.279      			$ & deg			\\
\hline
\vrule width 0pt height 10pt
$n_{\rm sky}$		& 344				\\
$n_{\rm ao}$		& 12600				\\
$n_{\rm lc}$		& 6852				\\
$n$			& 19796				\\
$\chi^2_{\rm sky}$	& 321				\\
$\chi^2_{\rm ao}$	& 36108				\\
$\chi^2_{\rm lc}$	& 52482				\\
$\chi^2$		& 59313				\\
\hline
\end{tabular}
\end{table}

\end{document}